\documentclass[aip,jcp,twocolumn,superscriptaddress]{revtex4}%
\usepackage{epsfig,amssymb,amsmath,amsthm,amsfonts,amsbsy,mathrsfs}
\usepackage{graphicx,color}
\usepackage{amsmath}
\usepackage{amssymb}%
\renewcommand{\Im}{\mathfrak{Im}}

\begin{document}
\title{Electronic Structure of Large-Scale Graphene Nanoflakes}

\author{Wei Hu}
%\thanks{Corresponding author. E-mail: gyrw4588@mail.ustc.edu.cn}
\thanks{Corresponding author. E-mail: whu@lbl.gov}
\affiliation{Computational Research Division, Lawrence Berkeley
National Laboratory, Berkeley, CA 94720, USA}

%\affiliation{Hefei National Laboratory for Physical Sciences at
%Microscale, and Department of Chemical Physics, University of
%Science and Technology of China, Hefei, Anhui 230026, China}

\author{Lin Lin}
\thanks{Corresponding author. E-mail: linlin@lbl.gov}
\affiliation{Department of Mathematics, University of California,
Berkeley, CA 94720, USA}
\affiliation{Computational Research Division, Lawrence Berkeley
National Laboratory, Berkeley, CA 94720, USA}

\author{Chao Yang}
\thanks{Corresponding author. E-mail: cyang@lbl.gov}
\affiliation{Computational Research Division, Lawrence Berkeley
National Laboratory, Berkeley, CA 94720, USA}

\author{Jinlong Yang}
\thanks{Corresponding author. E-mail: jlyang@ustc.edu.cn}
\affiliation{Hefei National Laboratory for Physical Sciences at
Microscale, and Department of Chemical Physics, University of
Science and Technology of China, Hefei, Anhui 230026, China}
\affiliation{Synergetic Innovation Center of Quantum Information and
Quantum Physics, University of Science and Technology of China,
Hefei, Anhui 230026, China}

\date{\today}

\pacs{ }

\begin{abstract}

With the help of the recently developed SIESTA-PEXSI method [J.
Phys.: Condens. Matter \textbf{26}, 305503 (2014)], we perform
Kohn-Sham density functional theory (DFT) calculations to study the
stability and electronic structure of hexagonal graphene nanoflakes
(GNFs) with up to 11,700 atoms. We find the electronic properties of
GNFs, including their cohesive energy, HOMO-LUMO energy gap, edge
states and aromaticity, depend sensitively on the type of edges
(ACGNFs and ZZGNFs), size and the number of electrons. We observe
that, due to the edge-induced strain effect in ACGNFs, large-scale
ACGNFs' cohesive energy decreases as their size increases. This
trend does not hold for ZZGNFs due to the presence of many edge
states in ZZGNFs. We find that the energy gaps $E_g$ of GNFs all
decay with respect to $1/L$, where $L$ is the size of the GNF, in a
linear fashion. But as their size increases, ZZGNFs exhibit more
localized edge states. We believe the presence of these states makes
their gap decrease more rapidly. In particular, when $L$ is larger
than 6.40 $nm$, we find that ZZGNFs exhibit metallic
characteristics. Furthermore, we find that the aromatic structures
of GNFs appear to depend only on whether the system has $4N$ or
$4N+2$ electrons, where $N$ is an integer.

%With the help of the recently developed pole expansion and selected
%inversion (PEXSI) method [J. Phys.: Condens. Matter \textbf{25},
%295501 (2013)], we perform Kohn-Sham density functional theory (DFT)
%calculations to study the stability and electronic structure of
%large-scale hexagonal graphene nanoflakes (GNFs) with thousands of
%atoms. Our calculations show that several quanities such as
%the cohesive energy and the HOMO-LUMO energy gap of a GNF depend
%sensitively on its size, edge type (i.e., arm chair vs. zigzag)
%and electron count.  For large GNFs, the cohesive
%energy of arm chair edged GNFs (ACGNFs) decreases rapidly as
%its size increases.  This is in sharp contrast with zigzag edged GNFs (ZZGNFs).
%Therefore, ACGNFs, especially large ACGNFs, are thermodynamically more stable
%and easier to produce experimentally than ZZGNFs. The HOMO-LUMO energy gap
%of GNFs decreases proportionaly to $1/L$, where $L$ is the diameter of the
%GNF. The rapid reduction of the HOMO-LUMO gap associated with ZZGNFs
%for large $L$ is due to the presence of many single electron states
%in which the electron density is concentrated along the edges of the ZZGNF.
%We refer to these states as edge states.
%These large ZZGNFs exibit metallic characteristics when their diameters
%are larger than 6.40 $nm$.  \CY{need something about why we care about
%aromaticity...}The aromaticity of GNFs is only determined
%by their electron count, which is either 4$N$ or 4$N$+2, where $N$ is
%an integer. The high stability of ACGNFs can be interpreted in terms of
%the edge-induced strain and the Clar's rule.

\end{abstract}

\maketitle

\section{Introduction}

Graphene, a two-dimensional (2D) sp$^2$-hybridized carbon sheet, has
recently received considerable interest owing to its outstanding
properties~\cite{Scinece_306_666_2004, NatureMater_6_183_2007,
ChemReV_107_718_2007}, such as its high carrier mobility, which is
important for graphene-based electronic devices, such as field
effect transistors (FETs). However, electronic devices fabricated
from graphene typically show a small on-off ratio due to its zero
bandgap. Therefore, many bandgap engineering techniques have been
developed both experimentally and theoretically to open a small band
gap in graphene~\cite{PRL_99_216802_2007, Nature_6_652_2007,
NatureMater_6_770_2007}. One of these techniques involves cutting 2D
graphene into finite-sized one-dimensional (1D) graphene nanoribbons
(GNRs)~\cite{PRL_98_206805_2007, Science_319_1229_2008,
JACS_130_4216_2008, Science_323_1701_2009, NaturePhysics_7_616_2011,
Nature_444_347_2006, NanoLett_6_2748_2006, PRL_97_216803_2006,
PRL_99_186801_2007, JACS_130_4224_2008, JACS_131_17728_2009} and
zero-dimensional (0D) graphene nanoflakes
(GNFs)~\cite{Science_320_356_2008, AdvFunctMater_18_3506_2008,
NatureMater_8_235_2009, PRB_81_085430_2010, PRB_82_045409_2010,
AdvMater_22_505_2010, Carbon_67_721_2014, JCP_140_074304_2014}.
Theoretically, significant efforts~\cite{Nature_444_347_2006,
NanoLett_6_2748_2006, PRL_97_216803_2006, PRL_99_186801_2007,
JACS_130_4224_2008, JACS_131_17728_2009} based on first-principles
calculations have been made to characterize properties of GNRs with
respect to the atomic configuration of their edges, which are of
either the armchair (AC) or zigzag (ZZ) types. These properties can
be used to guide bandgap engineering in 1D GNRs for graphene-based
electronic devices.

%Based on the
%tight-binding method and density functional theory, Son $et
%al.$\cite{PRL_97_216803_2006} have presented scaling rules for the
%band gaps $E_g$ of GNRs as a function of their widths ($L$ =
%3$N$/3$N$+1/3$N$+2, $N$ is an integer) and edges (ACGNRs and
%ZZGNRs). The tight-binding results show that ACGNRs with the widths
%$L$ = 3$N$+2 are metallic, but others ($L$ = 3$N$/3$N$+1) are all
%semiconducting, following a inversely relationship of $E_g$(3$N$)
%$\geq$ $E_g$(3$N$+1) $>$ $E_g$(3$N$+2) = 0 without considering the
%structural optimization of ACGNRs. But, first-principles
%calculations based on the density functional theory show that ACGNRs
%with different widths are all semiconducting, following a different
%relationship of $E_g$(3$N$+1) $>$ $E_g$(3$N$) $>$ $E_g$(3$N$+2) $>$
%0 due to the effect of ACGNRs' edge-induced strain. Such kind of
%rules are usefull for designing graphene-based electronic devisees,
%but the precise control over the widths of GNRs is still a technical
%challenge in experiments.

In this study, we focus on 0D GNFs, which are also known as graphene
quantum dots~\cite{PartPartSystCharact_31_415_2014}. Experimentally,
GNFs have been studied due to their unique properties and potential
applications~\cite{JACS_134_5718_2012, JPCC_116_5531_2012,
ACSNano_6_8203_2012, ACSNano_7_1239_2013}. In particular, for large
GNFs with lateral dimensions up to 20 $nm$, the dependency of the
electronic structure on the size and edge type was demonstrated by
experiments~\cite{NatureMater_8_235_2009}. It has been shown that
ZZGNFs exhibit metallic features and have localized edge states.
%It has been
%shown that when carbon atoms are arranged in a zigzag fashion along
%the edge of graphene, localized edge states exist, and they induce
%metallic features in ZZGNFs.
%

%Theoretically, the structural and electronic properties of small
%GNFs with hundreds of atoms have been widely studied with
%first-principles calculations~\cite{PRB_81_085430_2010,
%PRB_82_045409_2010, AdvMater_22_505_2010, Carbon_67_721_2014,
%JCP_140_074304_2014}. Their stability and the HOMO-LUMO energy
%gaps have been calculated.  However, for large GNFs, theoretical
%studies have been limited to simple H\"{u}ckel theory~\cite{JACS_109_3721_1987},
%pseudo-$\pi$ method~\cite{FaradayDiscuss_135_309_2007} and
%tight-binding method~\cite{PRB_77_235411_2008, PRB_82_155445_2010}.
%This is mainly due to the lack of computational tools that can be
%used to perform large-scale first-principles calculations that involve
%thousand or tens of thousands of atoms.

Theoretically, the structural and electronic properties of small
GNFs with up to hundreds of atoms have been studied with
first-principles calculations~\cite{PRB_81_085430_2010,
PRB_82_045409_2010, AdvMater_22_505_2010, Carbon_67_721_2014,
JCP_140_074304_2014}. The stability and the HOMO-LUMO energy
gaps have been calculated. However, for large GNFs, theoretical
studies have been limited. Most of the studies are based on the
H\"{u}ckel theory~\cite{JACS_109_3721_1987},
pseudo-$\pi$ method~\cite{FaradayDiscuss_135_309_2007} or
tight-binding method~\cite{PRB_77_235411_2008, PRB_82_155445_2010}.
This limitation is mainly due to the lack of computational tools that can be
used to perform large-scale first-principles calculations that involve
thousand or tens of thousands of atoms.

\begin{figure}[htbp]
\begin{center}
\includegraphics[width=0.5\textwidth]{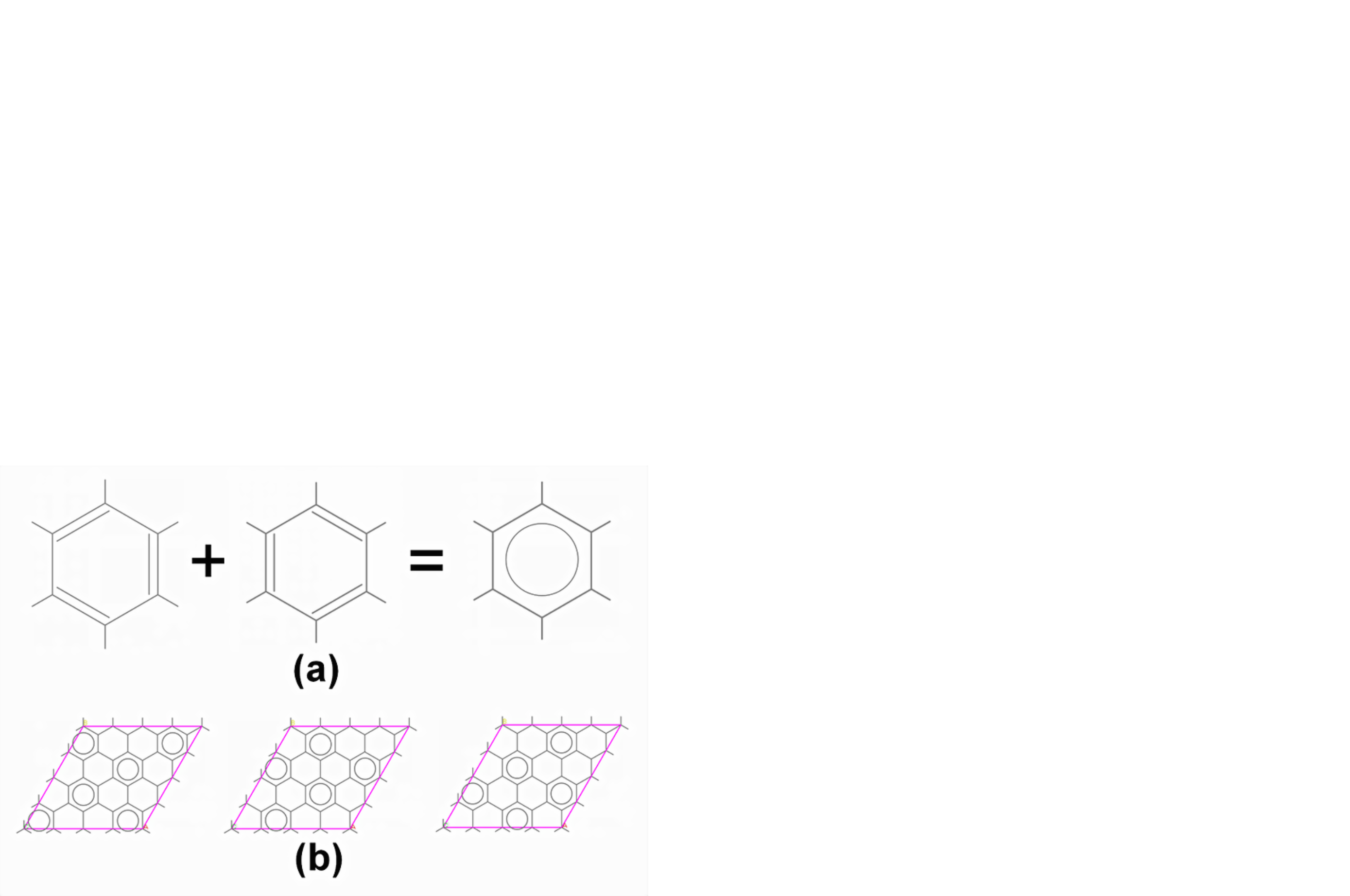}
\end{center}
\caption{(Color online) The Kekul\'{e} and Clar structure models for
(a) benzene and (b) graphene. Doublet and circle denote C=C double
bond and benzenoid ring, respectively.} \label{fig:Clar1}
\end{figure}

The hexagonal arrangement of carbon atoms in GNFs suggests that they
may share similar properties with other graphene based aromatic compounds
such as polycyclic aromatic hydrocarbons (PAHs)~\cite{JPC_56_311_1952,
ChemRev_101_1385_2001, ChemRev_101_1267_2001}, carbon nanotubes
(CNTs)~\cite{JOrgChem_69_4287_2004, OrgLett_9_4267_2007,
JPCC_113_862_2009} and GNRs~\cite{PRL_101_096402_2008,
JACS_132_3440_2010, SciRep_3_2030_2013}, whose
electronic structures can be characterized by their Kekul\'{e}
bonding structures, which contain alternating single and double
bonds within a hexagonal system such as those found
in a benzene molecule shown in Fig.~\ref{fig:Clar1}(a).
For benzene, there are two different Kekul\'{e} structures that are
distinguished by the locations of single and double bonds.
The resonance of these two complementary structures results in
what is known as a Clar sextet~\cite{Clar_1964, Clar_1972}.
For some polycyclic aromatic compounds such as graphene,
Clar sextets can appear at several possible locations. It follows
from Clar's theory that these sextets are disjoint and separated
by Kekul\'{e} structures. Different but equivalent Clar's formulas
can be derived based on the positions of the sextets. For example,
Fig.~\ref{fig:Clar1}(b) shows three Clar's formula for a
graphene~\cite{JACS_132_3440_2010}.
However, it is not clear whether Clar's theory remains valid for
large GNFs. First principle calculations based density functional theory
may be used to answer this question.

%The electronic structures of these
%systems can be interpreted in terms of what is called the Clar's rule~\cite{}.
%However, little work has been done on applying the Clar's rule
%on large-scale GNFs with thousands of atoms using results obtained from
%first-principles calculations.

%In the present work, we study the size and edge effects on the
%stability and electronic structure of large-scale hexagonal
%GNFs with thousands of atoms using first-principles density functional
%theory (DFT) based calculations. We compute the cohesive energies,
%structural and electronic properties of GNFs, and show that
%these quantities depend
%sensitively on the size and edge type of the GNFs of interest. We
%find that large-scale armchair-edged GNFs (ARGNFs) with thousands of
%atoms are more stable, and thus easier to assembled experimentally
%than zig-zag-edged GNFs (ZZGNFs). The HOMO-LUMO energy gap (denoted by $E_g$)
%of GNFs decreases proportionally to $1/L$, where $L$
%is the size of the GNF.  But $E_g$ associated with
%ZZGNFs decreases more rapidly as $L$ increases due to the presence of
%many edge states. There are different $\pi$-electron distributions for inner
%and outer states close to the Fermi level in GNFs, only depending
%the total number of electrons (independent of edge types), following
%a simple 4$N$/4$N$+2 rule ($N$ is an integer). \CY{need to clarify inner/outer/edge states, something wrong with the last sentence}

In this paper, we perform large scale first-principle calculations
for GNFs systems up to $11,700$ atoms, using the recently developed
PEXSI
method~\cite{LinLuYingE2009,LinLuYingEtAl2009,LinYangMezaEtAl2011,JPCM_25_295501_2013,JacquelinLinYang2014}
implemented in SIESTA~\cite{siestapexsi}. We report the computed
cohesive energies and HOMO-LUMO energy gaps for ACGNFs and ZZGNFs.
We predict that large ACGNFs are the most stable type of GNFs, thus
are easier to form than ZZGNFs in the experiments. The stability of
large ACGNFs can be understood by examining edge-induced strain for
ACGNFs of different sizes. We find that, as the system size
increases, the portion of ACGNF atoms that have small or zero strain
also increases. This trend renders large ACGNFs more stable than
small ACGNFs. We find that the HOMO-LUMO energy gap (denoted by
$E_g$) of ACGNFs and ZZGNFs decreases with respect to the system
size. Quantitatively, the relationship between $E_{g}$ and $L$ can
be described by $E_{g} = \alpha/L + \beta$, where $L$ is the size of
the GNF, and $\alpha$ and $\beta$ are some constants.  We find that
the HOMO-LUMO energy gap associated with ZZGNFs decreases more
rapidly than that associated with ACGNFs. We calculate the local
density of states (LDOS) and projected density of states (PDOS)
associated with the HOMO and LUMO states for both ACGNFs and ZZGNFs.
We find that the LDOS of ZZGNFs exhibits features that result from
increasingly significant contribution by the edges as the system
size increases, while the opposite holds for ACGNFs. By examining
the LDOS for the HOMO state, we identify two aromatic structures of
ACGNFs with different stability characteristics. In particular, we
find that the aromatic structure of ACGNFs depends on whether the
system has $4N$ or $4N+2$ electrons ($N$ is an integer), and the
induced stability character can be interpreted in terms of the
competition between Clar's theory for inner structure and the steric
effects of boundary structure in organic chemistry.

\section{Theoretical Models and Methods}

%In this work, we focus on hexagonal GNFs of different sizes.
%All GNFs we consider have diameter up to 20 $nm$.
%Both ACGNFs and ZZGNFs are examined. Each
%GNF contains either $4N$ or 4$N$+2 electrons, where $N$ is an
%integer, depending on its atomic configuration near the corner. For
%ACGNFs, the chemical formulas associated with these two different
%types of configurations are C$_{18n^2-30n+12}$H$_{12n-12}$ and
%C$_{18n^2-18n+6}$H$_{12n-6}$ respectively, where $n$ is also an
%integer. All ZZGNFs have similar structures but their width can
%differ. The chemical formulas of ZZGNFs with both 4$N$ and 4$N$+2
%electrons can be expressed by C$_{6n^2}$H$_{6n}$, for some integer
%$n$. A number of ACGNFs from C$_{42}$H$_{18}$ to
%C$_{11400}$H$_{300}$ and ZZGNFs from C$_{24}$H$_{12}$ to
%C$_{1014}$H$_{78}$ are considered in this study. The atomic
%geometries of some of these ACGNFs and ZZGNFs are shown in
%Fig.~\ref{fig:Structure}.

\begin{figure*}[htb]
\centering
\includegraphics[width=0.8\textwidth]{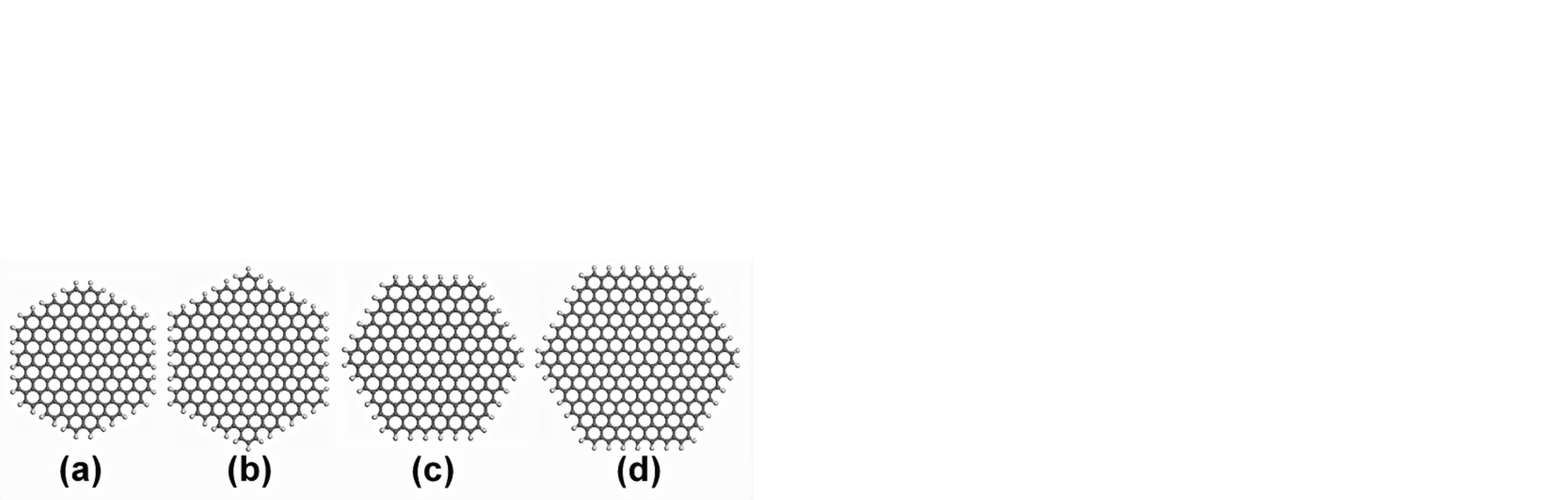}
\caption{(Color online) Atomic geometries of ACGNFs and ZZGNFs, (a)
C$_{180}$H$_{36}$ (4$N$), (b) C$_{222}$H$_{42}$ (4$N$+2), (c)
C$_{216}$H$_{36}$ (4$N$) and (d) C$_{294}$H$_{42}$ (4$N$+2). The
white and gray balls denote hydrogen and carbon atoms,
respectively.} \label{fig:Structure}
\end{figure*}

Both ACGNFs and ZZGNFs we consider have diameters below 20 $nm$.
Each GNF contains either 4$N$ or 4$N$+2 electrons ($N$ is an
integer), and the number of electrons depends on the atomic
configuration near the corners of the GNF. For ACGNFs, the
chemical formulae associated with these two different types of
configurations are C$_{18n^2-30n+12}$H$_{12n-12}$ and
C$_{18n^2-18n+6}$H$_{12n-6}$, respectively ($n$ is an integer). All
ZZGNFs share similar structures but have different widths. The
chemical formula of ZZGNFs with both $4N$ and $4N+2$ electrons can
be expressed by the same formula C$_{6n^2}$H$_{6n}$ ($n$ is an
integer). First-principle calculations for a number of ACGNFs from
C$_{42}$H$_{18}$ to C$_{11400}$H$_{300}$ and ZZGNFs from
C$_{24}$H$_{12}$ to C$_{1014}$H$_{78}$ are performed in this study.
The atomic geometries of some of these ACGNFs and ZZGNFs are shown
in Fig~\ref{fig:Structure}.

We use the Kohn-Sham DFT based electronic structure analysis
implemented in the SIESTA~\cite{PRB-1996-SIESTA} software package to
study properties of the GNFs discussed above. When performing DFT
calculations for these GNFs, we include 20 {\AA} vacuum space in
each of the $X$, $Y$ and $Z$ directions, which is sufficiently large
for separating the interactions between neighboring slabs. We choose
the PBE exchange correlation functional~\cite{PRL_77_3865_1996},
which generally gives a good description of electronic structures of
GNRs~\cite{JACS_132_3440_2010, SciRep_3_2030_2013} and
GNFs~\cite{JACS_134_5718_2012, JPCC_116_5531_2012}. We use the
double zeta plus polarization orbital basis set (DZP) to describe
the valence electrons within the framework of a linear combination
of numerical atomic orbitals (LCAO)~\cite{PRB-2001-LCAO}. All atomic
coordinates are fully relaxed using the conjugate gradient (CG)
algorithm until the energy and force convergence criteria of
10$^{-4}$ $eV$ and 0.04 $eV$/{\AA} respectively are reached. All
calculations are performed on the Edison system available at the
National Energy Research Scientific Computing (NERSC) center.

Due to the large number of atoms contained in the GNFs under study,
the standard diagonalization (DIAGON) method in SIESTA, which is
based on the ScaLAPACK~\cite{ScaLAPACK} software package, becomes prohibitively
expensive. Therefore, we use the recently developed pole expansion
and selected inversion (PEXSI)
technique~\cite{siestapexsi}
to reduce the computational time without sacrificing accuracy even
for metallic systems. The PEXSI technique allows the evaluation of
physical quantities such as electron density, energy, atomic force
to be performed without calculating any eigenvalue or eigenfunction. The
resulting SIESTA-PEXSI method can be highly scalable to more than
$10,000$ cores. It can effectively reduce the wall clock time.

To demonstrate the efficiency and accuracy of PEXSI for GNFs here,
we measure the average wall clock time spent in each self-consistent
field iteration for both PEXSI and diagonalization (DIAGON) methods
implemented in SIESTA for C$_{2382}$H$_{138}$. The PEXSI calculation
is performed using $40$ poles for all systems. We find that the time
used by DIAGON is 5 times more than that used by PEXSI when these calculations
were performed on 640 cores. The difference of total energy between
DIAGON and PEXSI calculations is less than 10$^{-4}$ $eV$ per atom.
The accuracy of PEXSI calculation can be further improved by simply
increasing the number of poles. The performance gain of PEXSI
relative to DIAGON becomes more substantial as the system size
increases. This is due to the $\mathcal{O}(N^{3/2})$ asymptotic
complexity of PEXSI for a quasi-2D system consisting of $N$ atoms,
which is superior to the $\mathcal{O}(N^{3})$ complexity of the
DIAGON method. Furthermore, the PEXSI method has much higher
parallel scalability than the DIAGON method in SIESTA when performed
on massively parallel computing platforms (with more than 1000
cores). As an example, we compare the wall clock time required to
perform one self-consistent field (SCF) iteration on
C$_{11400}$H$_{300}$. We found that the computational time required
by DIAGON is 23 times of that used by the PEXSI method in SIESTA
when the computation is performed on 2560 cores.

%{\color{blue} The following paragraph seems to be out of place. What does it
%have anything to do with what we present. Are we trying to verify
%Clar's rule? or are we just using structures predicted by
%Clar's rule to perform electronic structure calculation?
%I made an attemp to connect this with what's presented later,
%but it needs more work.}

%{\color{blue} We use SIESTA-PEXSI to compute both the total energy and
%various types of density of states (DOS). The standard
%DOS allows us to obtain the energy gap between
%the highest occupied molecular orbital (HOMO) and lowest
%unoccupied molecular orbital (LUMO). The distribution of
%electrons associated with these states can be deduced from
%the projected DOS (PDOS) and local DOS (LDOS). The SIESTA-PEXSI
%solver allows us to obtain DOS, PDOS and LDOS without
%diagonalizing the Kohn-Sham Hamiltonian. Hence it is
%extremely efficient for large GNFs that contain more than
%a thousand atoms. }

In the SIESTA-PEXSI solver, various types of density of states (DOS)
can be evaluated without computing any eigenvalue or eigenfunction as
well. The standard DOS allows us to obtain the energy gap between
the highest occupied molecular orbital (HOMO) and lowest unoccupied
molecular orbital (LUMO). The DOS can be computed via a procedure
called {\em inertia counting}, which is based on direct factorization
of sparse matrices and is described in detail in~\cite{siestapexsi}.
The inertia counting procedure can efficiently provide DOS at
arbitrary place along the spectrum with very high resolution, and
the DOS near the Fermi energy can be used to identify the energy levels of the
HOMO and LUMO state, and therefore the HOMO-LUMO band gap up to the
resolution of the DOS. For example, our calculated energy gap of
benzene (C$_6$H$_6$) is 5.25 $eV$ with the PEXSI method, in
agreement with the DIAGON method (5.24 $eV$) in SIESTA.

The spatial distribution of electrons associated with HOMO, LUMO, and
other states along the spectrum can be deduced from the local DOS
(LDOS), defined as
as
\[
\rho_L(r,\varepsilon) = 2 \sum_{i}\delta(\varepsilon-\varepsilon_i)|\psi_i(r)|^2,
\]
where $\psi_i$ is a Kohn-Sham orbital and $\varepsilon_i$ is the
corresponding Kohn-Sham energy. The LDOS $\rho(r,\varepsilon)$
provides an approximation to the electron density contributed by
electron states whose corresponding energies are near $\varepsilon$.
For instance, large values of LDOS on the edges of a GNF indicates
the presence of {\em edge states}, which are electronic states in which
relatively high electron density is found near the edges of
the GNF.  In the atomic orbital representation,
\[
\rho_L(r,\varepsilon) = 2 \sum_{i} \sum_{\mu,\nu} \varphi_{\mu}(r)
\varphi_{\nu}(r) \delta(\varepsilon-\varepsilon_i) c_{\mu,i} c_{\nu,i},
\]
where $\mu,\nu$ are atomic orbital indices, $\varphi_{\mu}$ is the
$\mu$th atomic orbital, $c_{\mu,i}$ is
the $\mu$th component of the $i$th Kohn-Sham eigenvector
$\psi_{i}$,
%which is the
%coefficient associated with the $\mu$th atomic orbital in the linear
%expansion of the eigenfunction in terms of atomic orbitals
and $H,S$ are the finite dimensional Hamiltonian and the overlap matrix
corresponding to the atomic orbitals, respectively.
For a given $\varepsilon$, $\rho_L(r,\varepsilon)$ can be computed
efficiently by the PEXSI method without diagonalizing the Kohn-Sham
Hamiltonian. The expression used in SIESTA-PEXSI is~\cite{siestapexsi}:
\[
  \rho_L(r,\varepsilon)\approx \frac{2}{\pi}
  \sum_{\mu,\nu}\varphi_{\mu}(r)\varphi_{\nu}(r)
    \Im \left[ H - (\varepsilon+i\eta) S\right]^{-1}_{\mu,\nu},
\]
where $\eta$ is a small broadening parameter describing the resolution
of the LDOS, and $\Im$ denotes the imaginary part. When $\varepsilon$
is chosen to be near the HOMO and LUMO energy level, the corresponding
LDOS provides accurate approximation of the electronic structure of
the HOMO and LUMO states.

The projected DOS (PDOS) measures the contribution of $\mu$th atomic
orbital (hence the atom itself) to the DOS around the energy level
$\varepsilon$. The definition of the PDOS is
\[
g_\mu(\varepsilon) = \frac{2}{N_a}\sum_i \sum_{\nu} c_{\nu,i} c_{\mu,i}S_{\mu,\nu}
\delta(\varepsilon-\varepsilon_i).
\]
Similar to the LDOS, the PDOS can be evaluated by the PEXSI technique
efficiently using the expression
\[
g_\mu(\varepsilon)\approx \frac{2}{N_{a} \pi}
  \sum_{\nu} S_{\mu,\nu} \Im \left[ H - (\varepsilon+i\eta) S\right]^{-1}_{\mu,\nu}.
\]
Such a procedure avoids the computation of eigenvalues and eigenfunctions
of the Kohn-Sham Hamiltonian, and is very efficient for large systems containing thousands of atoms.

\section{Results and Discussion}
%\CY{needs an introduction or summary}
%{\color{blue} In this section, we present computational results obtained from
%using SIESTA-PEXSI to solve the Kohn-Sham equations associated with both
%ACGNFs and ZZGNFs. We examine GNFs with both $4N$ and $4N+2$ electrons.
%We use these results to analyze the stability and electronic properties of
%GNFs.   We show that some of these results can be interpreted by existing
%theories on ...... We also make some new predictions for large scale GNFs....}

In this section, we present computational results obtained from
using SIESTA-PEXSI to study properties of GNFs. These properties
include the stability, HOMO-LUMO energy gap, the presence
of edge states and the aromatic structure.

\subsection{Stability}

\begin{figure}[htbp]
\begin{center}
\includegraphics[width=0.5\textwidth]{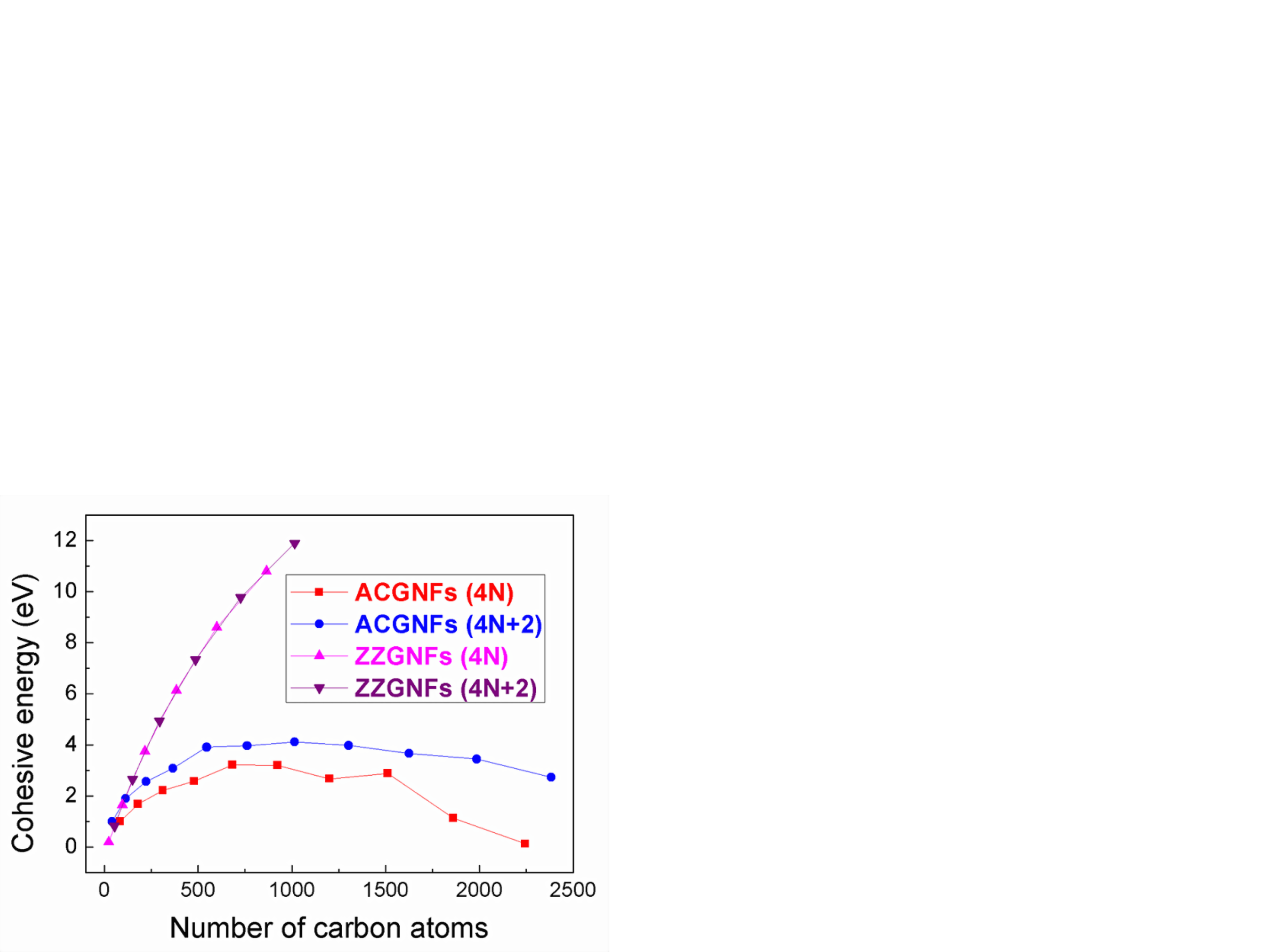}
\end{center}
\caption{(Color online) Cohesive energy $E_c$ ($eV$) of ACGNFs and
ZZGNFs with different total number of electrons (4$N$ and 4$N$+2,
$N$ is an integer) as a function of the number of carbon atoms.}
\label{fig:Ec}
\end{figure}

The stability of a GNF can be deduced from its cohesive energy, which is
defined as
\[
E_{c}=E_{GNF}-N_C\mu_C-N_H\mu_H,
\]
where $E_{GNF}$ represents the total energy of the GNF, $\mu$$_{C}$
and $\mu$$_{H}$ are the chemical potentials of carbon and hydrogen
atoms respectively, and $N_C$ and $N_H$ correspond to the number of
carbon and hydrogen atoms in the GNF respectively. Fig.~\ref{fig:Ec}
shows that the stability of GNFs depends strongly on their sizes and
edge types. For small GNFs with up to hundreds of atoms, the
cohesive energies of armchair edged GNFs (ACGNFs) and zigzag edged
GNFs (ACGNFs) all increase with respect to the number of carbon
atoms, which confirms the previous theoretical
results~\cite{Carbon_67_721_2014}. However, the rate of increase is
much higher for ZZGNFs than that for ACGNFs.

%We are interested in how the stability of a GNF varies with respect to
%its size (measured by its diameter $L$) and its edge type.
%Depending on how a GNF is cut out from a graphene sheet, its edge can
%be either of the arm chair (AC) type or of the zigzag (ZZ) type.

%Fig.~\ref{fig:Ec} \cite{Carbon_67_721_2014} \CY{why do we need a reference for
%the figure?} shows that the stability of GNFs depends strongly on
%their sizes and edge types.

%Interestingly, for large GNFs with thousands of atoms,
%Fig.~\ref{fig:Ec} shows that the cohesive energies of ACGNFs decrease
%with respect to the number of carbon atoms. This observation is very
%different from the trend observed for ZZGNFs, which exhibts a
%continued increase in cohesive energy as the number of atoms in the
%system increases. Therefore, we predict that large ACGNFs with
%thousands of atoms to be thermodynamically more stable and easier
%to produce experimentally than ZZGNFs. We also observe that
%large-scale ACGNFs with 4$N$ electrons are slightly more stable than
%ACGNFs with 4$N$+2 electrons.

For large GNFs with thousands of atoms, Fig.~\ref{fig:Ec} shows that the
cohesive energies of ACGNFs decrease with respect to the number of
carbon atoms. This observation is very different from the trend observed
for ZZGNFs, which exhibits a continued increase in cohesive energy as
the number of atoms in the system increases. Therefore, we predict that
large ACGNFs with thousands of atoms to be thermodynamically more stable
and easier to produce experimentally than ZZGNFs. We also observe that
large ACGNFs with 4$N$ electrons are slightly more stable than
ACGNFs with 4$N$+2 electrons.

The increased stability of large scale ACGNFs can be understood from
edge induced bond strain. Edge induced bond strain results from the
process of cutting graphene into hexagonal nanoflakes. The carbon
atoms in the outer layers of a nanoflake tends to relax and stretch
outward once they are cut away from the graphene. For GNRs, the
edge-induced strain has an important influence on their electronic
properties~\cite{JACS_132_3440_2010} such as the energy
gaps~\cite{PRL_97_216803_2006}. Here we find that edge-induced
strain also plays an important role on the stability of ACGNFs.
Edge-induced strain of a GNF can measured by
\[
\delta_{C-C}=(L_{GNFs}-L_G)/L_G,
\]
where $L_{GNFs}$ and $L_G$ represent the equilibrium carbon-carbon
bond length in a GNF and in an ideal monolayer of graphene,
respectively~\cite{PRL_97_216803_2006}. The carbon-carbon bond
length in an ideal monolayer of graphene is $L_G$ = 1.425
{\AA}~\cite{NatureMater_6_183_2007}.

\begin{figure*}[htb]
\centering
\includegraphics[width=0.8\textwidth]{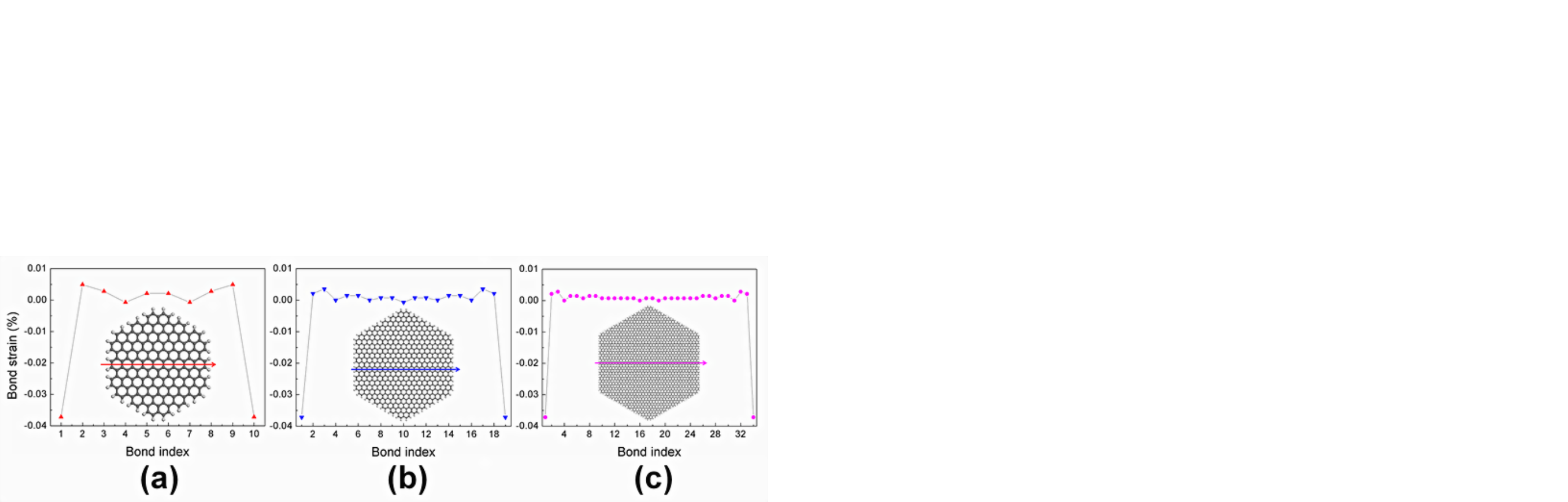}
\caption{(Color online) Edge-induced strain of carbon-carbon bonds
in ACGNFs, (a) C$_{180}$H$_{36}$, (b) C$_{684}$H$_{72}$ and (c)
C$_{2244}$H$_{132}$. Carbon-carbon bonds of ACGNFs marked along the
arrow direction are considered as shown in the insert.}
\label{fig:Strain}
\end{figure*}

We plot the computed rotationally averaged edge-induced strain
associated with carbon atoms in different layers of three ACGNFs
(C$_{180}$H$_{36}$, C$_{684}$H$_{72}$ and C$_{2244}$H$_{132}$) in
Fig.~\ref{fig:Strain}. The positive strain values indicate that the
C=C double bonds inside these ACGNFs are longer than the ideal bond
length in a graphene. The increased bond length renders ACGNFs less
stable compared to graphene. Furthermore, these carbon-carbon bonds
become even longer when they are closer to the edges of ACGNFs.
However, as the size of an ACGNF increases, its carbon-carbon bond
length becomes shorter. It eventually converges to that of an ideal
monolayer graphene. Therefore, the edge-induced strain in an ACGNF
is weakened as its size increases.  The weakened strain in large
ACGNFs (with more than a thousand atoms) makes them more stable
compared to small ACGNFs.

\subsection{HOMO-LUMO energy gap}

\begin{figure}[htbp]
\begin{center}
\includegraphics[width=0.5\textwidth]{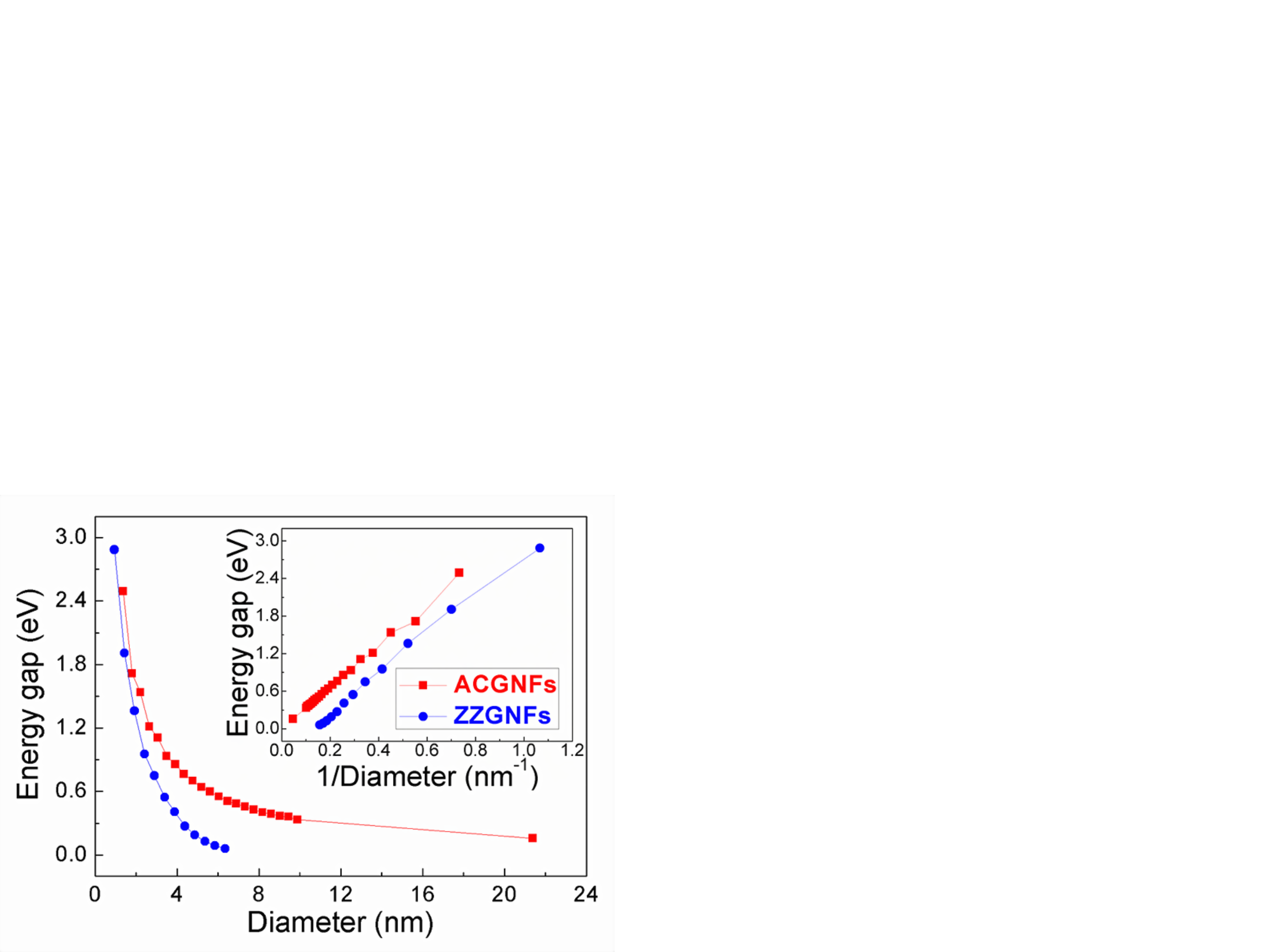}
\end{center}
\caption{(Color online) Energy gap $E_g$ (eV) of ACGNFs and ZZGNFs
as a function of sizes $L$ ($nm$) (Diameter). Energy gap
($E_g$)-size ($1/L$) relation of ACGNFs and ZZGNFs is shown in the
insert.} \label{fig:Eg}
\end{figure}

Fig.~\ref{fig:Eg} shows how the calculated HOMO-LUMO energy gaps $E_g$
($eV$) change with respect to the diameters $L$ ($nm$) of ACGNFs and
ZZGNFs, respectively. Our calculations show that the HOMO-LUMO gaps
of ACGNFs and ZZGNFs decrease as $L$ increases. The decrease in
$E_g$ can be attributed to the quantum confinement effect~\cite{PRL_90_037401_2003, JAP_108_094303_2010, CTC_1021_49_2013}. A
linear least squares fitting yields $E_g$ = 3.37/$L$ for ACGNFs and
$E_g$ = -0.62 + 3.97/$L$ for ZZGNFs respectively. These results are
close to previous models constructed from experimental measurements
($E_g$ = 1.57 $\pm $ 0.21/$L$$^{1.19\pm0.15}$)~\cite{NatureMater_8_235_2009} obtained from scanning tunneling
spectroscopy, and other theoretical predictions ($E_g$ =
1.68/$L$)~\cite{NatureMater_6_183_2007} based on quantum confinement
and the linear dispersion analysis of graphene.

%A linear relationship between $E_g$ and $1/L$ can be observed from the
%inset of Fig.~\ref{fig:Eg}.
%A linear least squares fitting yields $E_g$ = 3.37/$L$
%for ACGNFs and $E_g$ = -0.62 + 3.97/$L$ for ZZGNFs respectively.
%These results are close to previous models constructed from
%experimental measurements ($E_g$ = 1.57 $\pm $ 0.21/$L$$^{1.19\pm0.15}$)
%\cite{NatureMater_8_235_2009} obtained from scanning tunneling spectroscopy,
%and other theoretical predictions ($E_g$ = 1.68/$L$)~\cite{NatureMater_6_183_2007} based on quantum confinement and the linear dispersion analysis of
%graphene.

%We notice that the HOMO-LUMO energy gap of ZZGNFs decreases more rapidly with
%respect to $L$ than that of ACGNFs. This observation is consistent
%with previous analysis obtained from a tight-binding
%model~\cite{PRB_82_155445_2010}. The more rapid decrease in HOMO-LUMO energy
%gap is likely to be caused by the presence of edge states in which
%electron density is concentrated near the edge of a ZZGNF. We call
%these states {\em edge states}.
%Experimental studies\cite{NatureMater_8_235_2009} have shown that
%there are indeed many localized edge states with electron density
%concentrated on on carbon atoms within the zigzag edges of ZZGNFs,
%whereas no armchair edge state has been detected in
%ACGNFs.

We notice that the HOMO-LUMO energy gap of ZZGNFs decreases more rapidly
with respect to $L$ than that of ACGNFs. This observation is consistent
with previous analysis obtained from a tight-binding
model~\cite{PRB_82_155445_2010}. The more rapid decrease in HOMO-LUMO
energy gap is likely to be caused by the presence of edge states
whose electron densities concentrate near the edges of ZZGNFs.  Experimental
studies~\cite{NatureMater_8_235_2009} have shown that there are indeed
many localized edge states concentrating on carbon atoms along the
edges of ZZGNFs, whereas no edge state has been detected in ACGNFs.

%We also observe that large ZZGNFs with a diameter larger than 6.40 $nm$
%exhibit metallic features. This is in good agreement with the
%experimental measurements~\cite{NatureMater_8_235_2009}. On the other
%hand, ACGNFs are all found to be semiconducting with small energy gaps
%similar to ACGNRs~\cite{NanoLett_6_2748_2006, PRL_97_216803_2006,
%PRL_99_186801_2007}. However, the HOMO-LUMO gap of ACGNFs decreases
%monotonically as $L$ increase (i.e., $E_g$ $\propto$ $1/L$), whereas the
%HOMO-LUMO gaps of ACGNRs vary intricately with their widths, which is
%3$N$/3$N$+1/3$N$+2, where $N$ is an integer~\cite{PRL_97_216803_2006}.
%{\color{blue} Because large ACGNFs are energetically more stable, and
%because their HOMO-LUMO energy gaps can be easily controlled, we believe
%they are more suitable for graphene-based electronic devices than
%ACGNRs.}

We also observe that large ZZGNFs with a diameter larger than 6.40
$nm$ exhibit metallic features. This is in good agreement with the
experimental measurements~\cite{NatureMater_8_235_2009}. On the
other hand, ACGNFs are all found to be semiconducting with small
energy gaps similar to ACGNRs~\cite{NanoLett_6_2748_2006,
PRL_97_216803_2006, PRL_99_186801_2007}. However, the HOMO-LUMO gap
of ACGNFs decreases monotonically as $L$ increases (i.e., $E_g$
$\propto$ $1/L$), whereas the HOMO-LUMO gaps of ACGNRs intricately
depends on their widths (3$N$/3$N$+1/3$N$+2, where $N$ is an
integer)~\cite{PRL_97_216803_2006}.
Therefore, large ACGNFs show higher stability, and
their HOMO-LUMO energy gaps can be easily controlled for
graphene-based electronic devices.

\subsection{Edge states}

%As we indicated earlier that the small HOMO-LUMO gap of a GNF may be
%correlated with the presence of electron orbitals localized on the edges
%of the GNFs. We call electron states associated with these orbitals
%edge states.

\begin{figure*}[htb]
\centering
\includegraphics[width=0.75\textwidth]{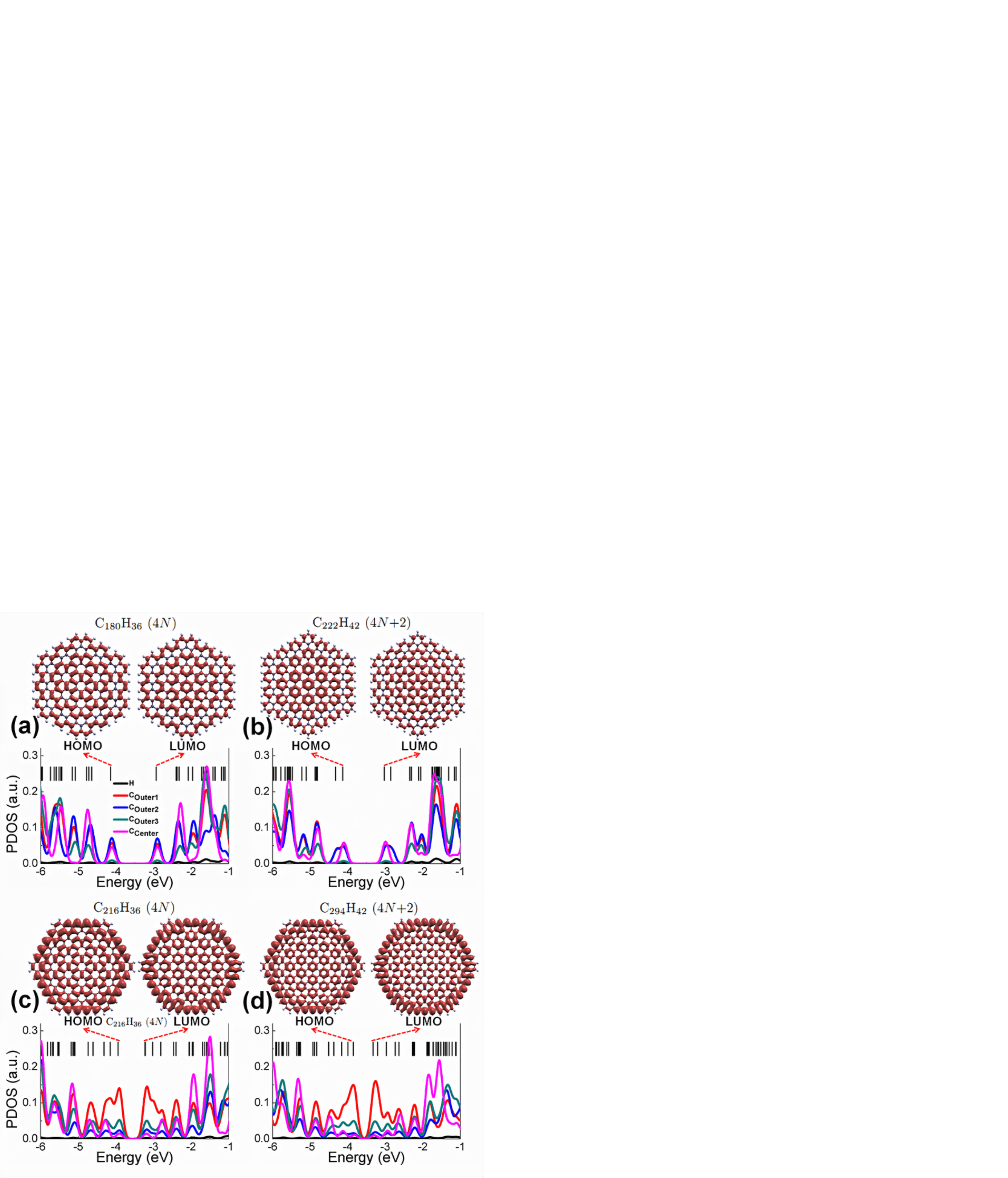}
\caption{(Color online) Energy levels and projected density of
states (PDOS) of small ACGNFs and ZZGNFs, (a) C$_{180}$H$_{36}$
(4$N$), (b) C$_{222}$H$_{42}$ (4$N$+2), (c) C$_{216}$H$_{36}$ (4$N$)
and (d) C$_{294}$H$_{42}$ (4$N$+2), including PDOS per atom of
hydrogen atoms (H), outermost (C$_{Outer1}$), second outer
(C$_{Outer2}$), third outer (C$_{Outer3}$) and central
(C$_{Center}$) carbon atoms. Their local density of states (LDOS) of
HOMO and LUMO are shown in the insert. Two kinds of delocalized
double bonds (CH=CH-C=C-CH=CH and CH-CH=C-C=CH-CH) in outer region
of HOMO states in ACGNFs are marked by pink arrows.}
\label{fig:PDOS1}
\end{figure*}

\begin{figure*}[htb]
\centering
\includegraphics[width=0.75\textwidth]{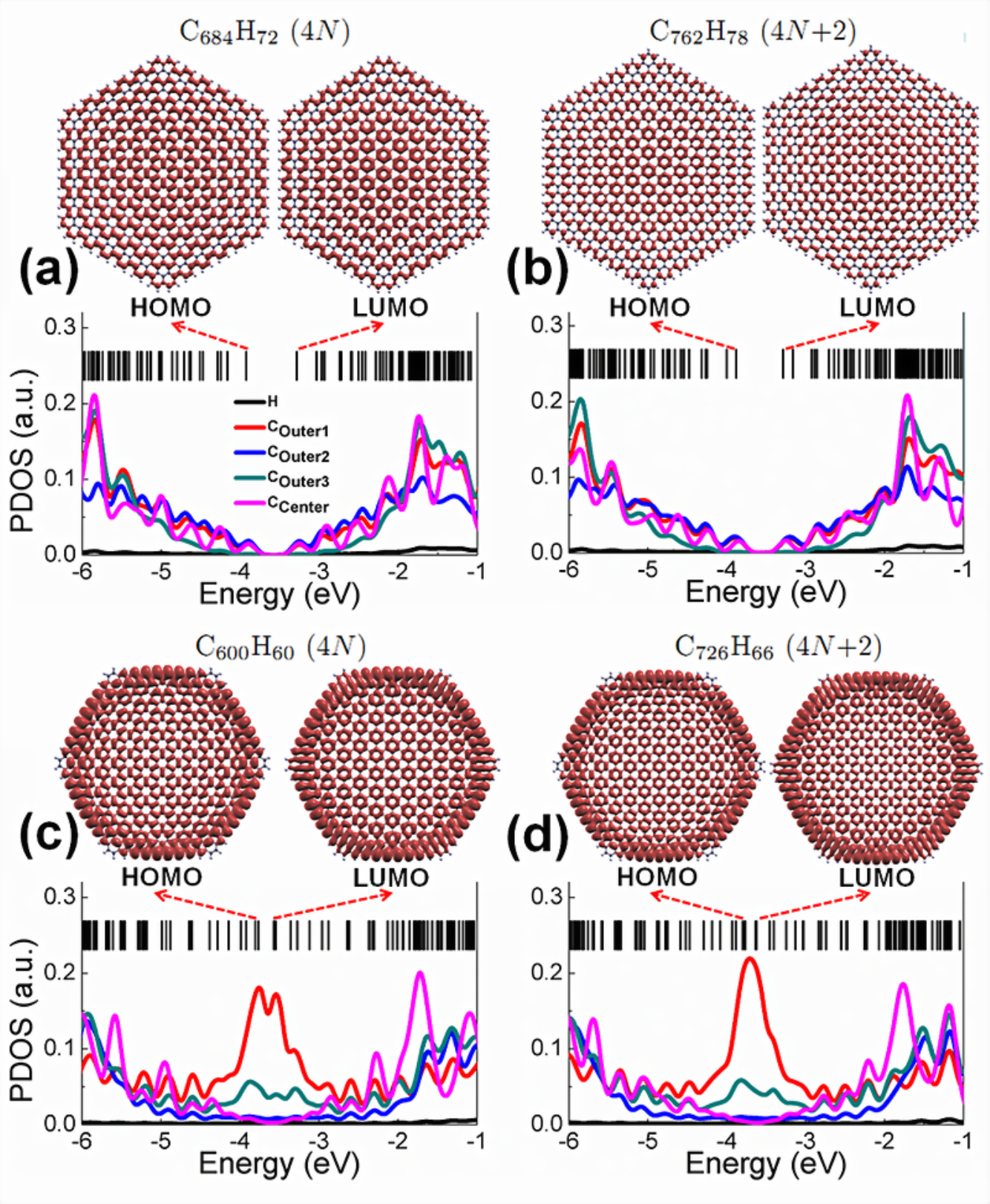}
\caption{(Color online) Energy levels and projected density of
states (PDOS) of large ACGNFs and ZZGNFs, (a) C$_{684}$H$_{72}$
(4$N$), (b) C$_{762}$H$_{78}$ (4$N$+2), (c) C$_{600}$H$_{60}$ (4$N$)
and (d) C$_{726}$H$_{66}$ (4$N$+2), including PDOS per atom of
hydrogen atoms (H), outermost (C$_{Outer1}$), second outer
(C$_{Outer2}$), third outer (C$_{Outer3}$) and central
(C$_{Center}$) carbon atoms. Their local density of states (LDOS) of
HOMO and LUMO are shown in the insert. Two kinds of delocalized
double bonds (CH=CH-C=C-CH=CH and CH-CH=C-C=CH-CH) in outer region
of HOMO states in ACGNFs are marked by pink arrows.}
\label{fig:PDOS2}
\end{figure*}

\begin{figure*}[htb]
\centering
\includegraphics[width=0.75\textwidth]{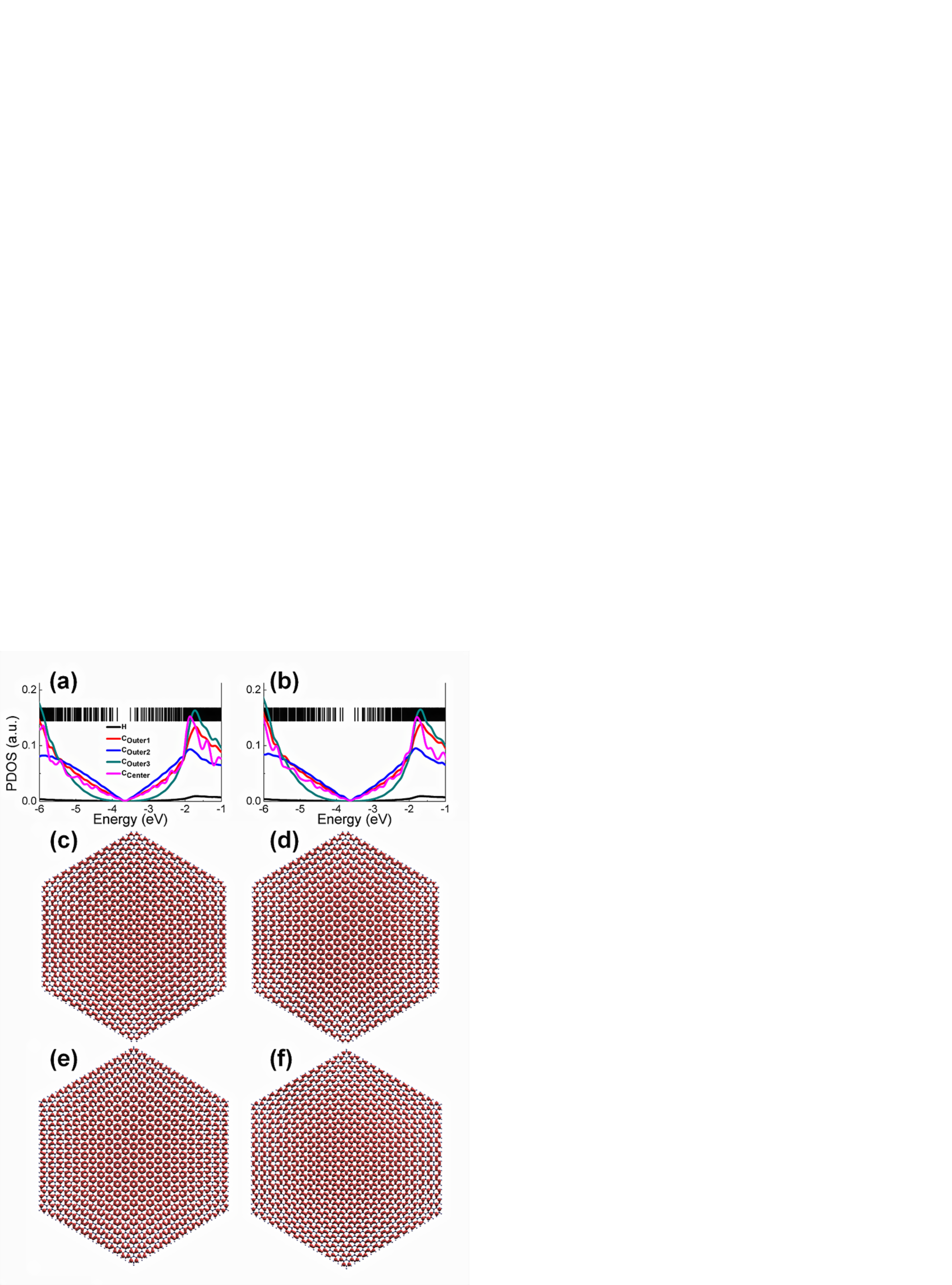}
\caption{(Color online) Energy levels and projected density of
states (PDOS) of large ACGNFs, (a) C$_{2244}$H$_{132}$ (4$N$), (b)
C$_{2382}$H$_{138}$ (4$N$+2), including PDOS per atom of hydrogen
atoms (H), outermost (C$_{Outer1}$), second outer (C$_{Outer2}$),
third outer (C$_{Outer3}$) and central (C$_{Center}$) carbon atoms,
and their corresponding local density of states (LDOS) of HOMO and
LUMO, (c) HOMO of C$_{2244}$H$_{132}$, (d) LUMO of
C$_{2244}$H$_{132}$, (e) HOMO of C$_{2382}$H$_{138}$ and (f) LUMO of
C$_{2382}$H$_{138}$.} \label{fig:PDOS3}
\end{figure*}

As we discussed earlier, the small HOMO-LUMO gap of a GNF is
related to the presence of edge states.  Edge states can be
revealed by computing local density of states (LDOS). In
Figs.~\ref{fig:PDOS1},~\ref{fig:PDOS2} and~\ref{fig:PDOS3}, we show
isosurfaces of the LDOS overlayed on atomic structures of various
GNFs for both $\varepsilon_{\mathrm{HOMO}}$ and
$\varepsilon_{\mathrm{LUMO}}$, where the HOMO and LUMO energies
$\varepsilon_{\mathrm{HOMO}}$ and $\varepsilon_{\mathrm{LUMO}}$ are
estimated from the DOS.

%defined as
%\[
%\rho_L(\varepsilon,r) = \sum_{i}\delta(\varepsilon-\varepsilon_i)|\psi_i(r)|^2,
%\]
%where $\psi_i$ is a Kohn-Sham orbital and $\varepsilon_i$ is the corresponding
%Kohn-Sham energy.  The LDOS $\rho(\varepsilon,r)$ provides an approximation
%to the electron density contributed by single electron states whose
%corresponding energies are near $\varepsilon$.
%For a given $\varepsilon$, $\rho_L(\varepsilon,r)$ can be computed
%efficiently using the PEXSI method without diagonalizing the Kohn-Sham
%Hamiltonian.  When an LDOS only has large values on the edges of a GNF, it
%indicates the presence of edge states.
%In Figs.~\ref{fig:PDOS1},~\ref{fig:PDOS2} and~\ref{fig:PDOS3}, we show LDOS isosurfaces overlayed on
%atomic structures of various GNFs for both $\varepsilon_{\mathrm{HOMO}}$ and
%$\varepsilon_{\mathrm{LUMO}}$, where the HOMO and LUMO energies
%$\varepsilon_{\mathrm{HOMO}}$ and $\varepsilon_{\mathrm{LUMO}}$ are
%estimated from the DOS, which can also be computed by SIESTA-PEXSI without
%diagonalizing the Kohn-Sham Hamiltonian.

We observe that for small ACGNFs shown in Fig.~\ref{fig:PDOS1}(a) and (b),
the LDOS plots associated with HOMO and LUMO states are not localized in any
particular region of the ACGNFs. However, for small ZZGNFs shown in
Fig.~\ref{fig:PDOS1}(c) and (d), the LDOS plots associated
with the HOMO and LUMO states show high levels of electron density
on the edges of these ZZGNFs. This is a clear indication
that edge states play an important role.

At the bottom of each subfigure in Fig.~\ref{fig:PDOS1}, we also
plot the projected DOS associated with atomic orbitals centered at
the hydrogen atoms and different layers of carbon atoms starting
from the outermost layer which forms the edge of the GNF. We can see
from Fig~\ref{fig:PDOS1}(a) and (b) that carbon atoms in the outer
layers and in the center of ACGNFs make equal contributions to the
HOMO and LUMO energy levels, although the contribution from the
third outer layer is significantly smaller than those from first two
outer layers. We also observe that hydrogen atoms make negligible
contribution.

Interestingly, the 3$n$th ($n$ is a small integer) outer layer
carbon atoms of ACGNFs, especially in large-scale, have no
contribution to their HOMOs and LUMOs. It can be seen from
circularly averaged hydrogen projected density of states (PDOS),
the outermost, the second and the third outer layer of carbon atoms
as well as carbon atoms at the center of the GNF and
their corresponding local HOMO and LUMO density of states (LDOS) as
shown in Figs.~\ref{fig:PDOS1},~\ref{fig:PDOS2} and~\ref{fig:PDOS3},
the third outer layer of carbon atoms in small ACGNFs (C$_{180}$H$_{36}$ and
C$_{222}$H$_{42}$) all have no contribution to their HOMOs and LUMOs
due to delocalized double bonds formed between the outermost and
second outer layer of carbon atoms. For large ACGNFs (C$_{684}$H$_{72}$,
C$_{762}$H$_{78}$, C$_{2244}$H$_{132}$ and C$_{2382}$H$_{138}$), the
third, sixth and even ninth outer carbon atoms also have no
contribution to their HOMOs and LUMOs. Furthermore, there is a
competition between the delocalized
double bonds near the armchair edge carbon atoms and
the all-benzenoid or non-benzenoid structure in the inner
region of the HOMO sates, as the sizes increase. But,
this effect does not exist in ZZGNFs, because rich outer edge states
dominate their HOMOs and LUMOs. Furthermore, the third outer layer of carbon
atoms even show more contribution to their HOMOs and LUMOs compared
with the second outer carbon atoms in ZZGNFs. Therefore, carbon
atoms in GNFs show different chemical activity. The difference depends on
their sizes, edge types and the total number of electrons.

Fig.~\ref{fig:PDOS2}(a) and (b) show that large ACGNFs, such as
C$_{684}$H$_{72}$ and C$_{762}$H$_{78}$ remain as semiconductors
with reduced HOMO-LUMO energy gaps. The estimated HOMO-LUMO gaps are
0.64 $eV$ for C$_{684}$H$_{72}$ and 0.60 $eV$ for C$_{762}$H$_{78}$.
The DOS and PDOS plots associated with these ACGNFs do not have
elevated peaks near the HOMO and LUMO levels. In contrast, large
ZZGNFs C$_{600}$H$_{60}$ and C$_{726}$H$_{66}$ exhibit metallic
characteristics, which can be seen from the much higher DOS values
near the Fermi level depicted in in Fig.~\ref{fig:PDOS2}(c) and (d).
The presence of a peak near the Fermi level is also correlated with
much higher LDOS values on the edges of the large ZZGNFs, which are
likely to be contributed by edge states, as we discussed earlier.
Furthermore, edge states in ZZGNFs become more localized as the
sizes increase, agreeing well with previous tight-binding
predictions~\cite{PRB_77_235411_2008}. Our results suggest that the
presence of many edge states tend to make large ZZGNFs
thermodynamically unstable, which is also manifested by their
relatively high cohesive energy levels compared with those of
ACGNFs. Similar conclusions have been reached for
GNRs~\cite{PRL_101_096402_2008, JACS_132_3440_2010,
SciRep_3_2030_2013}.

%We observe from the HOMO and LUMO LDOS plots shown in
%Figs.~\ref{fig:PDOS1} and~\ref{fig:PDOS2} that
%the $\pi$-electron distribution patterns in
%the inner region of ACGNFs and ZZGNFs are
%different from those in the outer regions of the nanoflakes.
%In the inner region, the HOMO and LUMO states of ACGNFs and ZZGNFs
%exhibit distinct aromatic and anti-aromatic characteristics.
%For example, in the inner region of C$_{180}$H$_{36}$, which is an ACGFN
%with $4N$ electrons , the carbon atoms appear to form $\pi$-electron Clar
%sextets in HOMO states shown in Fig~\ref{fig:PDOS1}(a), and
%empty hexagonal rings can be seen in the inner region
%of its LUMO states. Both the sextets and empty rings have a
%($\sqrt{3}$ $\times$ $\sqrt{3}$)$R$$30^{\circ}$ periodicity.
%These inner $\pi$-electron distribution
%patterns are similar to those found in GNRs~\cite{JACS_132_3440_2010}.

%In the outer region, alternating single and delocalized olefinic
%double bonds (CH-CH=C-C=CH-CH) can be observed in the HOMO and LUMO LDOS
%of ACGNFs, whereas the outer region of the ZZGNFs are dominated by edge states.

\subsection{Aromaticity}

%The aromatic structure in the inner region of an ACGNF depends on
%whether it has $4N$ or $4N+2$ electrons.
%ACGNFs with 4$N$+2 electrons belong to
%all-benzenoid polycyclic aromatic hydrocarbons with
%aromaticity~\cite{FaradayDiscuss_135_309_2007}. This type of atomic
%structure typically leads to high stability, high melting point, and
%low chemical reactivity~\cite{ChemReV_107_718_2007}, which is
%confirmed by our calculated cohesive energies.

We observe from the HOMO and LUMO LDOS plots shown in
Figs.~\ref{fig:PDOS1} and~\ref{fig:PDOS2} that
the $\pi$-electron distribution patterns in
the inner region of ACGNFs and ZZGNFs are
different from those in the outer regions of the nanoflakes.
In the inner region, the HOMO and LUMO states of ACGNFs and ZZGNFs
exhibit distinct aromatic and anti-aromatic characteristics.
For example, in the inner region of C$_{180}$H$_{36}$, which is an ACGNF
with $4N$ electrons, the carbon atoms appear to form $\pi$-electron Clar
sextets in HOMO states shown in Fig~\ref{fig:PDOS1}(a), and
empty hexagonal rings can be seen in the inner region
of its LUMO states. Both the sextets and empty rings have a
($\sqrt{3}$ $\times$ $\sqrt{3}$)$R$$30^{\circ}$ periodicity.
These inner $\pi$-electron distribution
patterns are similar to those found in GNRs~\cite{JACS_132_3440_2010}.

In the outer region, alternating single and delocalized olefinic
double bonds (CH-CH=C-C=CH-CH) can be observed in the HOMO and LUMO LDOS
of ACGNFs, whereas the outer region of the ZZGNFs are dominated by edge states.

The aromatic structure in the inner region of an ACGNF depends on
whether it has $4N$ or $4N+2$ electrons. ACGNFs with 4$N$+2
electrons belong to all-benzenoid polycyclic aromatic hydrocarbons
with aromaticity~\cite{FaradayDiscuss_135_309_2007}. We can observe
from Fig.~\ref{fig:PDOS1} (b) that this type of ACGNFs have unique
Clar formulas that correspond to all-benzenoid structures with a
($\sqrt{3}$ $\times$ $\sqrt{3}$)$R$$30^{\circ}$ periodicity. A
similar observation is made in~\cite{PRB_77_235411_2008} based on a
tight-binding model. However, ACGNFs with 4$N$ electrons appears to
have non-benzenoid structures that contain empty hexagonal rings
with a ($\sqrt{3}$ $\times$ $\sqrt{3}$)$R$$30^{\circ}$ periodicity
in the inner region of their HOMO states shown in
Fig.~\ref{fig:PDOS1}(a).

As we mentioned earlier, the outer regions of ACGNFs consist of
alternating single and delocalized olefinic double bonds formed
along the armchair edges.  However, the locations of the double
bonds are different for ACGNFs with 4N electron and those with 4N+2
electrons. For ACGNFs with 4N electrons, the bonding pattern can be
labeled by CH=CH-C=C-CH=CH, whereas for ACGNFs with 4N+2 electrons,
the pattern becomes CH-CH=C-C=CH-CH.Thus, the locations of chemical
addition reaction associated with ACGNFs with $4N$ electrons are
different from those associated with ACGNFs with $4N+2$ electrons.
Such difference may affect the carrier mobility along the edges of
ACGNFs, similar to the effects observed for
ACGNRs~\cite{JACS_131_17728_2009} as well as the stability of the
GNF.

%inner region of their HOMO states shown in Fig.~\ref{fig:PDOS1}(a).

%As we mentioned earlier, the outer regions of ACGNFs consist of
%alternating single and delocalized olefinic double bonds formed
%along the armchair edges.  However, the locations of the double
%bonds are different for ACGNFs with 4N electron and those with 4N+2 electrons.
%For ACGNFs with 4N electrons, the bonding pattern can be
%labeled by CH=CH-C=C-CH=CH, whereas for ACGNFs with 4N+2
%electrons, the pattern becomes CH=CH-C=C-CH=CH.

\begin{figure*}[htb]
\centering
\includegraphics[width=0.8\textwidth]{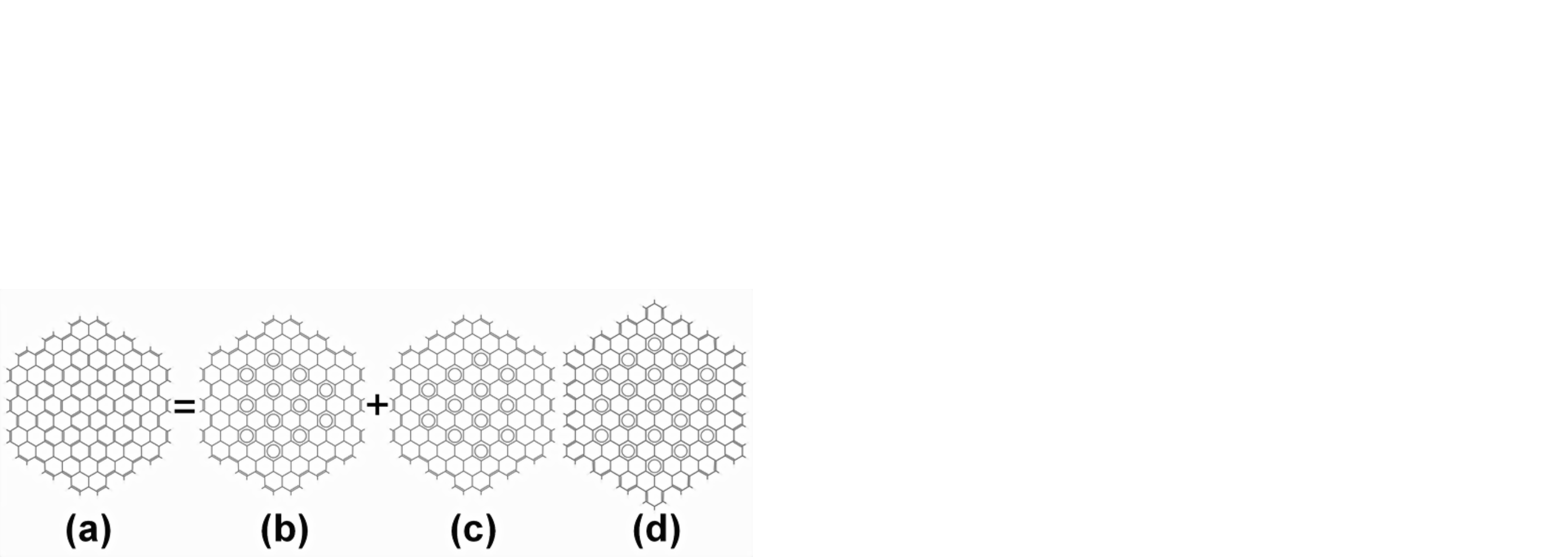}
\caption{(Color online) The Kekul\'{e} and Clar structure models of
ACGNFs, (a) the Kekul\'{e} formulas, (b) and (c) two Clar formulas
of C$_{180}$H$_{36}$ (4$N$), and (d) unique Clar formulas of
C$_{222}$H$_{42}$ (4$N$+2).} \label{fig:Clar2}
\end{figure*}

Based on the cohesive energy results we presented earlier, we predict
that ACGNFs with 4$N$ electrons are slightly more stable than
ACGNFs with 4$N$+2 electrons of comparable sizes.
This is particularly true for large ACGNFs with thousands of atoms.
However, the HOMO LDOS plot shows that ACGNFs with 4$N$ electrons
exhibits non-benzenoid structures that contain empty hexagonal
rings with a ($\sqrt{3}$ $\times$ $\sqrt{3}$)$R$$30^{\circ}$ periodicity,
which can be can be interpreted as a linear combination of two
Clar formulas in the inner region. Such a linear combination tends
to be less stable than all-benzenoid polycyclic aromatic hydrocarbons
(PAHs) with unique Clar formulas observed in ACGNFs with 4$N$+2
electrons as illustrated in Fig.~\ref{fig:Clar2}. However, the relative
stability of a system is determined both by the inner structure and by the
boundary structure, which is reflected here by the steric effects of the
$\pi$-electrons near the boundary and the locations of delocalized
olefinic double bonds.  The slightly higher stability of ACGNFs with
4$N$ electrons compared to 4$N$+2 electrons indicates the competition
between Clar's theory for the inner structure and the steric effects of
the boundary structure.

For ZZGNFs, we also observe non-benzenoid or all-benzenoid
structures in the inner regions of their HOMO states depending on
whether they have $4N$ or $4N+2$ electrons.  However, the
difference in these inner region aromatic structures appears
to have little effect on their cohesive energy.  This also indicates the
importance of the boundary, which is dominated by the edge states
(\.{C}H-C-\.{C}H-C-\.{C}H-C) for ZZGNFs.

\section{Summary and Conclusions}

\begin{table*}
  \caption{The PEXSI method calculated stability and electronic
properties (energy gap, inner and outer HOMO) of large GNFs
with different edges (ACGNFs and ZZGNFs) and total number of
electrons (4$N$/4$N$+2, $N$ is an integer).}  \label{multiprogram}
\begin{tabular}{ccccccc} \\ \hline \hline
GNFs & & \multicolumn{2}{c}{ACGNFs} & & \multicolumn{2}{c}{ZZGNFs} \ \\
Number of electrons & & 4$N$ & 4$N$+2 & & 4$N$ & 4$N$+2 \ \\
\hline
Stability & & Most stable & Stable & & Unstable & Unstable \ \\
          & &  \multicolumn{2}{c}{Low chemical reactivity} & & \multicolumn{2}{c}{High chemical reactivity} \ \\
Energy gap & & \multicolumn{2}{c}{$E_g$ = 3.37/$L$} & & \multicolumn{2}{c}{$E_g$ = -0.62 + 3.97/$L$} \ \\
           & & \multicolumn{2}{c}{All semiconducting} && \multicolumn{2}{c}{Metallic for $L$ $>$ 6.40 $nm$} \\
Inner HOMO & & Non-benzenoid & All-benzenoid & & Non-benzenoid & All-benzenoid \ \\
           & & Two Clar formulas & Unique Clar formulas & & Two Clar formulas & Unique Clar formulas \ \\
Outer HOMO & & \multicolumn{2}{c}{Delocalized double bonds} & & \multicolumn{2}{c}{Rich localized edge states} \ \\
           & & CH=CH-C=C-CH=CH & CH-CH=C-C=CH-CH & & \multicolumn{2}{c}{\.{C}H-C-\.{C}H-C-\.{C}H-C}   \ \\
\hline \hline
\end{tabular}
\end{table*}

In summary, we investigate the effects of the sizes and edges on the
stability and electronic structure of hexagonal graphene nanoflakes
(GNFs) using first-principle calculations at an unprecedented scale.
This is enabled through the recently developed SIESTA-PEXSI method
for efficient treatment of large scale electronic structure
calculations even for systems of metallic characteristics.  The main
findings of this paper is given in Table~\ref{multiprogram},
summarizing the properties of ACGNFs and ZZGNFs with $4N$ and $4N+2$
electrons, respectively. The results presented in this study are
important to the understanding of size and edge dependency of GNFs
with potential applications for graphene-based electronic
applications at nanoscale.

\section{Acknowledgments}

This work is partially supported by National Key Basic Research
Program (2011CB921404, 2012CB922001), by NSFC (21121003, 91021004,
20933006, 11004180), by Strategic Priority Research Program of CAS
(XDB01020300). This work is also partially supported by the
Scientific Discovery through Advanced Computing (SciDAC) Program
funded by U.S. Department of Energy, Office of Science, Advanced
Scientific Computing Research and Basic Energy Sciences (W. H., L.
L. and C. Y.), and by the Center for Applied Mathematics for Energy
Research Applications (CAMERA), which is a partnership between Basic
Energy Sciences and Advanced Scientific Computing Research at the
U.S Department of Energy (L. L. and C. Y.). We thank the National
Energy Research Scientific Computing (NERSC) center, and the
USTCSCC, SC-CAS, Tianjin, and Shanghai Supercomputer Centers for the
computational resources.


\begin{thebibliography}{38}



\bibitem{Scinece_306_666_2004}K. S. Novoselov, A. K. Geim, S. V. Morozov, D. Jiang, Y. Zhang, S. V. Dubonos, I. V. Grigorieva, and A. A. Firsov, Scinece \textbf{306}, 666 (2004).

%Electric Field Effect in Atomically Thin Carbon Films


\bibitem{NatureMater_6_183_2007}A. K. Geim and K. S. Novoselov, Nature Mater. \textbf{6}, 183 (2007).

%The rise of graphene


\bibitem{ChemReV_107_718_2007}J. Wu, W. Pisula, and K. M\"{u}llen, Chem. Rev. \textbf{107}, 718 (2007).

%Graphenes as Potential Material for Electronics









\bibitem{PRL_99_216802_2007}E. V. Castro, K. S. Novoselov, S. V. Morozov, N. M. R. Peres, J. M.
B. Lopes dos Santos, J. Nilsson, F. Guinea, A. K. Geim, and A. H. C.
Neto, Phys. Rev. Lett. \textbf{99}, 216802 (2007).

%Biased Bilayer Graphene: Semiconductor with a Gap Tunable by the Electric Field Effect


%\bibitem{ACSNano_2_2301_2008}Z. H. Ni, T. Yu, Y. H. Lu, Y. Y. Wang, Y. P. Feng, and Z. X. Shen, ACS Nano \textbf{2}, 2301 (2008).

%Uniaxial Strain on Graphene: Raman Spectroscopy Study and Band-Gap Opening


%\bibitem{NaturePhysics_6_30_2010}F. Guinea, M. I. Katsnelson, and A. K. Geim, Nature Physics \textbf{6}, 30 (2010).

%Energy gaps and a zero-field quantum Hall effect in graphene by strain engineering






\bibitem{Nature_6_652_2007}F. Schedin, A. K. Geim, S. V. Morozov, E. W. Hill, P. Blake, M. I.
Katsnelson, and K. S. Novoselov, Nature Mater. \textbf{6}, 652
(2007).

%Detection of individual gas molecules adsorbed on graphene



\bibitem{NatureMater_6_770_2007}S. Y. Zhou, G.-H. Gweon, A. V. Fedorov, P. N. First, W. A. de Heer, D.-H. Lee, F. Guinea, A. H. C. Neto, and A. Lanzara, Nature Mater. \textbf{6}, 770 (2007).

%Substrate-induced bandgap opening in epitaxial graphene


%\bibitem{Nature_5_722_2010}C. R. Dean, A. F. Young, I. Meric, C. Lee, L. Wang, S. Sorgenfrei, K. Watanabe, T. Taniguchi, P. Kim, K. L. Shepard, and J. Hone, Nature Nanotech. \textbf{5}, 722 (2010).

%Boron nitride substrates for high-quality graphene electronics




%\bibitem{JCP_138_054701_2013}W. Hu, Z. Li, and J. Yang, J. Chem. Phys. \textbf{138}, 054701 (2013).

%Diamond as an inert substrate of graphene


%\bibitem{JCP_138_124706_2013}W. Hu, Z. Li, and J. Yang, J. Chem. Phys. \textbf{138}, 124706 (2013).

%Electronic and optical properties of graphene and graphitic ZnO nanocomposite structures


%\bibitem{JCP_139_154704_2013}W. Hu, Z. Li, and J. Yang, J. Chem. Phys. \textbf{139}, 154704 (2013).

%Structural, electronic, and optical properties of hybrid silicene and graphene nanocomposite














%\bibitem{Nature_454_319_2008}J. C. Meyer, C. O. Girit, M. F. Crommie, and A. Zett, Nature \textbf{454}, 319 (2008).

%Imaging and dynamics of light atoms and molecules on graphene


%\bibitem{NanoLett_8_173_2008}T. O. Wehling, K. S. Novoselov, S. V. Morozov, E. E. Vdovin, M. I. Katsnelson, A. K. Geim, and A. I. Lichtenstein, Nano Lett. \textbf{8}, 173 (2008).

%Molecular Doping of Graphene


%\bibitem{PRL_101_086402_2008}S. Y. Zhou, D. A. Siegel, A. V. Fedorov, and A. Lanzara, Phys. Rev. Lett. \textbf{101}, 086402 (2008).

%Metal to Insulator Transition in Epitaxial Graphene Induced by Molecular Doping












\bibitem{PRL_98_206805_2007}M. Y. Han, B. \"{O}zyilmaz, Y. Zhang, and P. Kim, Phys. Rev. Lett. \textbf{98}, 206805 (2007)

%Energy Band-Gap Engineering of Graphene Nanoribbons


\bibitem{Science_319_1229_2008}X. Li, X. Wang, L. Zhang, S. Lee, and H. Dai, Science \textbf{319}, 1229 (2008).

%Chemically Derived, Ultrasmooth Graphene Nanoribbon Semiconductors


\bibitem{JACS_130_4216_2008}X. Yang, X. Dou, A. Rouhanipour, L. Zhi , H. J. R\"{a}der, and K. M\"{u}llen, J. Am. Chem. Soc. \textbf{130}, 4216 (2008).

%Two-Dimensional Graphene Nanoribbons


\bibitem{Science_323_1701_2009}X. Jia, M. Hofmann, V. Meunier, B. G. Sumpter, J. Campos-Delgado.,
J. M. Romo-Herrera, H. Son., Y.-P. Hsieh, A. Reina, J. Kong, M.
Terrones, and M. S. Dresselhaus, Science \textbf{323}, 1701 (2009).

%Controlled Formation of Sharp Zigzag and Armchair Edges in Graphitic Nanoribbons


\bibitem{NaturePhysics_7_616_2011}C. Tao, L. Jiao, O. V. Yazyev, Y.-C. Chen, J. Feng, X. Zhang, R. B. Capaz, J. M. Tour, A. Zettl, S. G. Louie, H. Dai, and M. F. Crommie, Nature Physics \textbf{7}, 616 (2011).

%Spatially resolving edge states of chiral graphene nanoribbons








\bibitem{Nature_444_347_2006}Y.-W. Son, M. L. Cohen, and S. G. Louie, Nature \textbf{444}, 347 (2006).

%Half-metallic graphene nanoribbons


\bibitem{NanoLett_6_2748_2006}V. Barone, O. Hod, and G. E. Scuseria, Nano Lett. \textbf{6}, 2748 (2006).

%Electronic Structure and Stability of Semiconducting Graphene Nanoribbons


\bibitem{PRL_97_216803_2006}Y.-W. Son, M. L. Cohen, and S. G. Louie, Phys. Rev. Lett. \textbf{97}, 216803 (2006).

%Energy Gaps in Graphene Nanoribbons


\bibitem{PRL_99_186801_2007}L. Yang, C.-H. Park, Y.-W. Son, M. L. Cohen, and S. G. Louie, Phys. Rev. Lett. \textbf{99}, 186801
(2007).

%Quasiparticle Energies and Band Gaps in Graphene Nanoribbons


\bibitem{JACS_130_4224_2008}E. Kan, Z. Li, J. Yang, and J. G. Hou, J. Am. Chem. Soc. \textbf{130}, 4224 (2008).

%Half-Metallicity in Edge-Modified Zigzag Graphene Nanoribbons


\bibitem{JACS_131_17728_2009}M.-Q. Long, L. Tang, D. Wang, L. Wang, and Z. Shuai, J. Am. Chem. Soc. \textbf{131}, 17728 (2009).

%Theoretical Predictions of Size-Dependent Carrier Mobility and Polarity in Graphene




\bibitem{Science_320_356_2008}L. A. Ponomarenko, F. Schedin, M. I. Katsnelson, R. Yang, E. W. Hill, K. S. Novoselov, and A. K. Geim, Science \textbf{320}, 356 (2008).

%Chaotic Dirac Billiard in Graphene Quantum Dots


\bibitem{AdvFunctMater_18_3506_2008}N. G. Shang, P. Papakonstantinou, M. McMullan, M.
Chu, A. Stamboulis, A. Potenza, S. S. Dhesi, and H. Marchetto, Adv.
Funct. Mater. \textbf{18}, 3506 (2008).

%Catalyst-Free Efficient Growth, Orientation and Biosensing Properties of Multilayer Graphene Nanoflake Films with Sharp Edge Planes


\bibitem{NatureMater_8_235_2009}K. A. Ritter and J. W. Lyding, Nature Mater. \textbf{8}, 235 (2009).

%The influence of edge structure on the electronic properties of graphene quantum dots and nanoribbons











\bibitem{PRB_81_085430_2010}A. Kuc, T. Heine, and G. Seifert, Phys. Rev. B \textbf{81}, 085430 (2010).

%Structural and electronic properties of graphene nanoflakes


\bibitem{PRB_82_045409_2010}M. Wimmer, A. R. Akhmerov, and F. Guinea, Phys. Rev. B \textbf{82}, 045409 (2010).

%Robustness of edge states in graphene quantum dots


\bibitem{AdvMater_22_505_2010}G. Eda, Y.-Y. Lin, C. Mattevi, H. Yamaguchi, H.-A. Chen, I-S. Chen,
C.-W. Chen, and M. Chhowalla, Adv. Mater. \textbf{22}, 505 (2010).

%Blue Photoluminescence from Chemically Derived Graphene Oxide



\bibitem{Carbon_67_721_2014}N. Wohner, P. Lam, and K. Sattler, Carbon \textbf{67}, 721 (2014).

%Energetic stability of graphene nanoflakes and nanocones


\bibitem{JCP_140_074304_2014}S. K. Singh, M. Neek-Amal, and F. M. Peeters, J. Chem. Phys. \textbf{140}, 074304 (2014).

%Electronic proper ties of graphene nano-flakes: Energy gap, permanent dipole , termination e ffect, a nd Raman s pectroscopy





\bibitem{PartPartSystCharact_31_415_2014}M. Bacon, S. J. Bradley, and T. Nann, Part. Part. Syst. Charact. \textbf{31}, 415 (2014).

%Graphene Quantum Dots








\bibitem{JACS_134_5718_2012}E. Kan, W. Hu, C. Xiao, R. Lu, K. Deng, J. Yang, and H. Su, J. Am. Chem. Soc. \textbf{134}, 5718 (2012).

%Half-Metallicity in Organic Single Porous Sheets


\bibitem{JPCC_116_5531_2012}Y. Zhou, Z. Wang, P. Yang, X. Sun, X. Zu, and F. Gao, J. Phys. Chem. C  \textbf{116}, 5531 (2012).

%Hydrogenated Graphene Nanoflakes: Semiconductor to Half-Metal Transition and Remarkable Large Magnetism


\bibitem{ACSNano_6_8203_2012}S. Kim, S. W. Hwang, M.-K. Kim, D. Y. Shin, D. H. Shin, C. O. Kim,
S. B. Yang, J. H. Park, E. Hwang, S.-H. Choi, G. Ko, S. Sim, C.
Sone, H. J. Choi, S. Bae, and B. H. Hong, ACS Nano \textbf{6}, 8203
(2012).

%Anomalous Behaviors of Visible Luminescence from Graphene Quantum Dots: Interplay between Size and Shape


\bibitem{ACSNano_7_1239_2013}S. H. Jin, D. H. Kim, G. H. Jun, S. H. Hong, and S. Jeon, ACS Nano \textbf{7}, 1239 (2013).

%Tuning the Photoluminescence of Graphene Quantum Dots through the Charge Transfer Effect of Functional Groups


%\bibitem{Nanoscale_5_4015_2013}L. Li, G. Wu, G. Yang, J. Peng, J. Zhao, and J.-J. Zhu, Nanoscale \textbf{5}, 4015 (2013).

%Focusing on luminescent graphene quantum dots: current status and future perspectives







\bibitem{JACS_109_3721_1987}S. E. Stein and R. L. Brown, J. Am. Chem. Soc. \textbf{109}, 3721 (1987).

%p-Electron Properties of Large Condensed  Polyaromatic Hydrocarbons


\bibitem{FaradayDiscuss_135_309_2007}E. Steiner, P. W. Fowler, A. Soncini, and L. W. Jenneskens, Faraday Discuss. \textbf{135}, 309 (2007).

%Current-density maps as probes of aromaticity: Global and Clar p ring currents in totally resonant polycyclic aromatic hydrocarbons



\bibitem{PRB_77_235411_2008}Z. Z. Zhang, K. Chang, and F. M. Peeters, Phys. Rev. B \textbf{77}, 235411 (2008).

%Tuning of energy levels and optical properties of graphene quantum dots


\bibitem{PRB_82_155445_2010}A. D. G\"{u}\c{c}l\"{u}, P. Potasz, and P. Hawrylak, Phys. Rev. B \textbf{82}, 155445 (2010).

%Excitonic absorption in gate-controlled graphene quantum dots






\bibitem{JPC_56_311_1952}C. A. Coulson, J. Phys. Chem. \textbf{56}, 311 (1952).


\bibitem{ChemRev_101_1385_2001}T. M. Krygowski  and M. K. Cyra\'{n}ski, Chem. Rev. \textbf{101}, 1385 (2001).

%Structural Aspects of Aromaticity


\bibitem{ChemRev_101_1267_2001}M. D. Watson, A. Fechtenk\"{o}tter, and K. M\"{u}llen, Chem. Rev. \textbf{101}, 1267 (2001).

%Big Is Beautiful-¡°Aromaticity¡± Revisited from the Viewpoint of Macromolecular and Supramolecular Benzene Chemistry









\bibitem{JOrgChem_69_4287_2004}J. L. Ormsby and B. T. King, J. Org. Chem. \textbf{69}, 4287 (2004).

%Clar valence bond representation of p-bonding in carbon nanotubes


\bibitem{OrgLett_9_4267_2007}M. Baldoni, A. Sgamellotti, and F. Mercuri, Org. Lett. \textbf{9}, 4267 (2007).

%Finite-length models of carbon nanotubes based on clar sextet theory


\bibitem{JPCC_113_862_2009}M. Baldoni, D. Selli, A. Sgamellotti, and F. Mercuri, J. Phys. Chem. C \textbf{113}, 862 (2009).

%Electronic properties and stability of graphene nanoribbons: An interpretation based on Clar sextet theory












\bibitem{PRL_101_096402_2008}T. Wassmann, A. P. Seitsonen, A. M. Saitta, M. Lazzeri, and F. Mauri, Phys. Rev. Lett. \textbf{101}, 096402 (2008).

%Structure, Stability, Edge States, and Aromaticity of Graphene Ribbons


\bibitem{JACS_132_3440_2010}T. Wassmann, A. P. Seitsonen, A. M. Saitta, M. Lazzeri, and F. Mauri, J. Am. Chem. Soc. \textbf{132}, 3440 (2010).

%Clar's Theory, p-Electron Distribution, and Geometry of Graphene Nanoribbons


\bibitem{SciRep_3_2030_2013}Y. Li, Z. Zhou, C. R. Cabrera, and Z. Chen, Sci. Rep. \textbf{3}, 2030 (2013).

%Preserving the Edge Magnetism of Zigzag Graphene Nanoribbons by Ethylene Termination: Insight by Clar's Rule







\bibitem{Clar_1964}E. Clar, \emph{Polycyclic Hydrogarbons} (Academic Press, London, 1964).


\bibitem{Clar_1972}E. Clar, \emph{The Aromatic Sextet} (Wiley, New York, 1972).




\bibitem{LinLuYingE2009} L. Lin, J. Lu, L. Ying and W. E, Chin. Ann. Math. \textbf{30B}, 729 (2009).


\bibitem{LinLuYingEtAl2009} L. Lin, J. Lu, L. Ying, R. Car and W. E, Commun. Math. Sci. \textbf{7}, 755 (2009).


\bibitem{LinYangMezaEtAl2011} L. Lin, C. Yang, J. Meza, J. Lu, L. Ying and W. E, ACM Trans. Math. Software \textbf{37}, 40 (2011).


\bibitem{JPCM_25_295501_2013}L. Lin, M. Chen, C. Yang, and L. He, J. Phys.: Condens. Matter \textbf{25}, 295501 (2013).

%Accelerating atomic orbital-based electronic structure calculation via pole expansion and selected inversion


\bibitem{JacquelinLinYang2014} M. Jacquelin, L. Lin and C. Yang, arXiv:1404.0447


\bibitem{siestapexsi} L. Lin, A. Garc\'{\i}a, G. Huhs and C. Yang, J. Phys.: Condens. Matter \textbf{26}, 305503 (2014).

% SIESTA-PEXSI: massively parallel method for efficient and accurate ab initio materials simulation without matrix diagonalization






\bibitem{PRB-1996-SIESTA}P. Ordej\'{o}n, E. Artacho, and J. M. Soler, Phys. Rev. B \textbf{53}, R10441
(1996).

%Self-consistent order-N density-functional calculations for very large systems





\bibitem{PRL_77_3865_1996}J. P. Perdew, K. Burke, and M. Ernzerhof, Phys. Rev. Lett. \textbf{77}, 3865 (1996).

%Generalized Gradient Approximation Made Simple

\bibitem{PRB-2001-LCAO}J. Junquera, \'{O}. Paz, D. S\'{a}nchez-Portal, and E. Artacho, Phys.
Rev. B \textbf{64}, 235111 (2001).

%Numerical atomic orbitals for linear-scaling calculations



\bibitem{ScaLAPACK}T. Auckenthaler, V. Blum, H. J. Bungartz, T. Huckle, R. Johanni, L. Kr\"{a}mer, B. Lang, H. Lederer, and P. R. Willems, Parallel Comput. \textbf{37}, 783 (2011).




%\bibitem{PRB_51_1456_1995}P. Ordej\'{o}n, D. A. Drabold, R. M. Martin, and M. P. Grumbach, Phys. Rev. B \textbf{51}, 1456 (1995).

%Linear system-size methods for electronic-structure calculations


\bibitem{PRL_90_037401_2003}J. Raty, G. Galli, and T. van Buuren, Phys. Rev. Lett. \textbf{90}, 037401 (2003).

%Quantum confinement and fullerenelike surface reconstructions in nanodiamonds



\bibitem{JAP_108_094303_2010}J. Jiang, L. Sun, B. Gao, Z. Wu, W. Lu, J. Yang, and Y. Luo, J. Appl. Phys. \textbf{108}, 094303 (2010).

%Structure dependent quantum confinement effect in hydrogen-terminated nanodiamond clusters


\bibitem{CTC_1021_49_2013}W. Hu, Z. Li, and J. Yang, Comput. Theor. Chem. \textbf{1021}, 49 (2013).

%Surface and size effects on the charge state of NV center in nanodiamonds



\end{thebibliography}
\end{document}